\documentclass[a4paper,11pt]{article}
\pdfoutput=1 

\usepackage{jheppub}
\usepackage{color} 
\usepackage{float}
\usepackage{ulem}
\usepackage{braket}
\usepackage{dsfont}
\usepackage{subcaption}
\usepackage{float}
\usepackage{amsmath,graphicx,amssymb,bbm}
\usepackage[all]{xy}
\newcommand{\R}{\mathbb{R}}

\begin{document}

\title{Electromagnetic form factors of vector mesons in Einstein-dilaton holographic QCD}

\author[a]{Alfonso Ballon-Bayona,}
\author[b]{Tobias Frederico,}
\author[c,d]{Luis A. H. Mamani,}
\author[a]{Ad\~ao S. da Silva Junior,}
\author[b]{and Wayne de Paula}
\affiliation[a]{Instituto de F\'{i}sica, Universidade
Federal do Rio de Janeiro, \\
Caixa Postal 68528, RJ 21941-972, Brazil.}    
\affiliation[b]{Instituto Tecnol\'ogico de Aeron\'autica, DCTA, 12228-900 S\~ao Jos\'e dos Campos, Brazil}
\affiliation[c]{Centro de Ci\^encias Exatas e Tecnol\'ogicas,
Universidade Federal do Rec\^oncavo da Bahia,\\
Rua Rui Barbosa, 710, 44380-000, Cruz das Almas, Bahia, Brazil}
\affiliation[d]{Laborat\'orio de Astrof\'{\i}sica Te\'orica e Observacional,\\ 
Departamento de Ci\^encias Exatas e Tecnol\'ogicas,
Universidade Estadual de Santa Cruz,\\ 45650-000, Ilh\'eus, Bahia, Brazil}
\emailAdd{aballonb@if.ufrj.br}
\emailAdd{tobias@ita.br}
\emailAdd{luis.mamani@ufrb.edu.br}
\emailAdd{adaossj@pos.if.ufrj.br}
\emailAdd{wayne@ita.br}

\abstract{We investigate the electromagnetic form factors of the $\rho$ meson family within a holographic QCD framework based on Einstein-dilaton gravity with confining backgrounds. We obtain the elastic and transition form factors 
using a spectral decomposition of the three-point isospin current correlator and an independent  approach based on the Kaluza-Klein expansion. Furthermore, we establish a holographic dictionary of the current matrix element. The elastic form factors are in agreement with available lattice QCD results, whereas the transition form factors, not yet determined within lattice QCD, are predictions of the model. For the elastic case, the  electric form factor for the $\rho$ meson family exhibits a zero at $q^{2} = 6\, m_{\rho^{n}}^{2}$ and in particular for the ground state $q^{2} =  3.61\,\text{GeV}^{2}$, consistent with light-front and Dyson-Schwinger calculations. We also compute strong couplings between vector meson states and find some approximate selection rules. In the elastic case and low-$q^{2}$ regime, we calculate dimensionless electric radius, magnetic and quadrupole moments and compare with other approaches. In the high-$q^{2}$ regime, the form factors follow the expected scaling behavior, and the model predicts their asymptotic normalization. The results in the high-$q^{2}$ regime lead to  superconvergence relations and establish a new sum rule involving masses, decay constants and couplings.
}

\maketitle

\flushbottom

\section{Introduction}

The interaction of an external photon with hadronic matter serves as a precise probe of the internal structure of hadrons, providing a valuable tool for exploring the nonperturbative dynamics of QCD \cite{Weise:1974bk}. In the low-energy regime, this interaction can be studied through electromagnetic form factors, which are quantities that describe the spatial distribution of charges and magnetic moments inside the hadron and allow us to probe the size of the hadron through the electric, magnetic, and quadrupole radii. In this context, the $\rho$ meson plays an important role due to the vector meson dominance proposed by Sakurai \cite{Sakurai:1960ju, sakurai1969currents} from gauge non-abelian theories, according to which the electromagnetic interaction between a hadron and a photon is mediated by the $\rho$ meson so that the electromagnetic form factor of any hadron contains the $\rho$ meson pole.

The electromagnetic form factors of vector mesons have been investigated using several theoretical approaches, such as the Bethe-Salpeter equations \cite{Munz:1994ty}, Dyson-Schwinger methods \cite{Xu:2024vkn}, light front quantum field theory \cite{DeMelo:2018bim}, Nambu-Jona-Lasinio model \cite{Zhang:2022zim, Zhang:2024nxl}, and Lattice QCD calculations \cite{Owen:2015gva}. The Dyson-Schwinger equations constitute a set of coupled integral equations of the Green's functions of quantum field theory (see review \cite{Roberts:1994dr,Maris:2003vk}). This approach is commonly used in conjunction with the Bethe-Salpeter models \cite{Hernandez-Pinto:2024kwg}. Light-front quantum field theory is formulated onto the null-plane hypersurface, where the dynamical evolution is defined along the light-front time \cite{Brodsky:2011sk,dePaula:2026gtb,Marinho:2026auy}. The Nambu–Jona-Lasinio model is an effective non-renormalizable field theoretical approach that describes spontaneous chiral symmetry breaking.  However, a unified description of both elastic and transition form factors of vector mesons  following the approaches described above remains incomplete. One of the main challenges is the description of excited states of the $\rho$ meson and the strong couplings between them. 

The gauge/gravity duality \cite{Maldacena:1997re, Witten:1998qj, Gubser:1998bc}, also known as holography,  provides a theoretical framework to study strongly coupled gauge theories. This type of duality enables a rich interplay between field theory and gravity, grounded in insights from string theory, with numerous applications across both low-energy physics, such as condensed matter, and high-energy physics, including particle physics. In its original formulation, type IIB string theory on $AdS_{5}\times S^{5}$  is conjectured to be dual to $\mathcal{N}=4$ Super Yang-Mills theory in $\R^{1,3}$. This is known as the $AdS_5/CFT_4$ correspondence, where $AdS_5$ stands for five dimensional anti-de Sitter space and $CFT_4$ stands for a four dimensional conformal field theory.   

The application of gauge/gravity duality models to QCD is usually known as holographic QCD or AdS/QCD. The electromagnetic form factors of vector mesons have already been explored in phenomenological AdS/QCD models, such as the hard-wall model \cite{Grigoryan:2007vg} and soft-wall model \cite{Grigoryan:2007my}. 
In models involving D-brane systems, such as the Sakai-Sugimoto model and the Kuperstein-Sonnenschein model, the electromagnetic form factors of vector mesons have also been studied \cite{BallonBayona:2009ar,Bayona:2010bg}. In both AdS/QCD models and D-brane models, the electromagnetic current decomposes into vector meson states (a generalization of vector meson dominance), allowing the description of electromagnetic form factors in terms of strong couplings between vector meson states.  The fundamental state and the excited states of the $\rho$ meson are described by Sturm-Liouville modes or Kaluza-Klein modes associated with the extra dimension.

An important class of AdS/QCD models for conformal symmetry breaking and confinement was proposed in \cite{Gursoy:2007cb,Gursoy:2007er}, see also \cite{dePaula:2008fp,Li:2013oda,Ballon-Bayona:2017sxa}. These approaches are based on five dimensional Einstein-dilaton gravity and we will refer to it as Einstein-dilaton holographic QCD. This class of models incorporate a dynamical dilaton field consistent with Einstein's equations in five dimensions and lead to a deformation of the  anti-de Sitter ($AdS_5$) spacetime consistent with the confinement criterion \cite{Kinar:1998vq}. The four dimensional dual theory is conformal in the ultraviolet regime and conformal symmetry breaking leads to asymptotically linear spectra for hadronic states such as mesons, glueballs and nucleons. Moreover, it can be shown that the spectral decomposition of the hadronic correlators is consistent with QCD in the large $N_c$ limit \cite{Ballon-Bayona:2024yuz}. This constitutes an important advantage of this class of holographic models compared to others, such as the hard-wall model, which is confining but does not produce linear Regge trajectories, or the soft-wall model, which yields an exactly linear spectrum for vector mesons but does not satisfy the confinement criterion. Moreover, Eistein-dilaton holographic QCD presents advantages over holographic models involving D-branes, discussed above, since the latter usually contain Kaluza-Klein modes not present in QCD.  It is important to mention that there is another holographic QCD approach based on Einstein-dilaton gravity whose initial goal is not to reproduce linear confinement and the hadronic spectrum at zero temperature but instead to build a dilaton potential that reproduces the finite temperature equation of state of the quark-gluon plasma in $N_c=3$ lattice QCD  \cite{Gubser:2008ny,Gubser:2008yx}, see also \cite{DeWolfe:2010he,Rougemont:2023gfz}.

Einstein-dilaton holographic QCD has been successfully applied in hadron phenomenology to describe mass spectra, decay constants of vector mesons and nucleons \cite{dePaula:2009za, Ballon-Bayona:2024yuz}, glueball spectra and the trace anomaly \cite{Gursoy:2007er, Ballon-Bayona:2017sxa} as well as soft pomeron phenomenology \cite{Ballon-Bayona:2015wra,Ballon-Bayona:2017vlm}. Several extensions of this framework have been developed. Einstein-Maxwell-dilaton models are important to incorporate finite quark density \cite{Ballon-Bayona:2020xls} as well as inverse magnetic catalysis \cite{Rodrigues:2017iqi, Dudal:2021jav, Bohra:2020qom, Zhu:2023aaq}. The Einstein-tachyon-dilaton models describe the spontaneous breaking of chiral symmetry \cite{Ballon-Bayona:2023zal}, and the Einstein-axion-dilaton models describe the CP-violation in QCD \cite{Gursoy:2007er, Hamada:2020phg}. 

In this work, we investigate the electromagnetic form factors of the $\rho$ meson family using Einstein-dilaton holographic QCD following two approaches: the first one based on the explicit calculation of the three-point correlation function of isospin currents following the AdS/CFT dictionary and the other using the four-dimensional effective action obtained from the Kaluza-Klein expansion. In the first approach, we expand the action to cubic order in the gauge fields and express it in terms of the bulk to boundary propagators. Using the AdS/CFT dictionary we obtain the three-point correlator of isospin currents with the dynamical part depending on the propagators. Then we perform a spectral decomposition on two of the propagators to obtain the generalized form factor, which includes the elastic and transition form factors. 
The second approach starts with the five-dimensional action and, by applying the Kaluza-Klein expansion, obtains an effective four-dimensional action. Using the Feynman rules for the vertices and propagators in this action we obtain the matrix element of the isospin current and find the electromagnetic form factors, in complete agreement with the first approach. We then numerically evaluate the elastic and transition form factors, the strong couplings between vector mesons, and provide a full description of the electric, magnetic, and quadrupole form factors.

This paper is organized as follows: In section \ref{Sec:VMHQCD}, we present the Einstein-dilaton models that are used throughout the work, briefly reviewing the confinement criterion and conformal symmetry breaking in this class of models. We also present the five-dimensional action for the vector meson sector, the field equations, the mass spectrum and decay constants. Section \ref{Sec:FormFactors} is the most important part of this paper, where we calculate the three point correlator of isospin currents, perform the spectral decomposition, obtain the elastic and transition form factors, the strong couplings between vector mesons and provide a full description of the electric, magnetic, and quadrupole form factors. In section \ref{asympgeneralizedformfactors}, we provide a detailed description of the asymptotic behavior of elastic and transition form factors in the soft-wall and Einstein-dilaton models and from this analysis we obtain important sum rules for the couplings. Finally, in section \ref{section4}, we present our conclusions. In appendix \ref{App:kkexpansion}, we present the electromagnetic form factors of vector mesons derived from the Kaluza–Klein expansion method. In appendix \ref{vecsoftwall}, we provide a review of the electromagnetic form factors of vector mesons within the soft-wall model, which includes novel analytical results for the strong couplings and an exact selection rule. In appendix \ref{vecbtbp}, we present a numerical analysis of the bulk to boundary propagator.


\section{Vector mesons in Einstein-dilaton holographic QCD}\label{Sec:VMHQCD}

In this section we review Einstein-dilaton holographic QCD, focusing on the two analytical models considered in  \cite{Ballon-Bayona:2024yuz} and describe the dynamics of the vector meson sector, including a summary of the results for the spectrum and decay constants.  In section \ref{Sec:FormFactors} we will investigate the cubic couplings of the model and present a systematic description of the electromagnetic form factors of vector mesons in Einstein-dilaton holographic QCD. 

\subsection{Einstein-dilaton holographic QCD}\label{einsteindilatonmodels}

As described in the introduction, we are interested in holographic QCD models based on Einstein-dilaton gravity that are consistent with confinement and asymptotically linear spectrum. We start with the 5d Einstein-dilaton action in the Einstein frame
\begin{align}
    S = M_{p}^{3}N_{c}^{2} \int d^{5}x\, \sqrt{-g}\left[R - \dfrac{4}{3}\partial^{m}\Phi\partial_{m}\Phi + \ell^{-2}V(\Phi)\right],\label{edaction}
\end{align}
where $M_{p}^{3}$ is the Planck mass and $N_{c}$ the number of colors. The action \ref{edaction} includes the Ricci scalar $R$, the dilaton field, $\Phi$, the dilaton potential $V(\Phi)$ and the $AdS_{5}$ radius $\ell$. In this work, we will fix the $AdS_{5}$ radius as $\ell = 1$. By varying the action with respect to the metric, we obtain the Einstein-dilaton field equations
\begin{align}
    G_{mn} &=\dfrac{1}{2 M_{p}^{3}N_{c}^{2}}T_{mn},\\
    \nabla^{2}\Phi &+ \dfrac{3}{8}\dfrac{dV}{d\Phi} = 0,\,
\end{align}
where the Einstein tensor, $G_{mn}$, and energy-momentum tensor, $T_{mn}$, are respectively given by
\begin{align}
    G_{mn} = R_{mn} - \dfrac{R}{2}g_{mn} \quad , \quad 
    T_{mn} = M_{p}^{3}N_{c}^{2}\left[\dfrac{8}{3}\partial_{m}\Phi\partial_{n}\Phi + g_{mn}\mathcal{L}_{\Phi}\right],
\end{align}
and the dilaton lagrangian density is
\begin{align}
    \mathcal{L}_{\Phi} = -\dfrac{4}{3}g^{mn}\partial_{m}\Phi\partial_{n}\Phi + V(\Phi)\,.
\end{align}
The background usually used in holographic QCD models is given by the line element $5d$
\begin{align}
    ds^{2} =\dfrac{1}{\zeta(u)^{2}}\left[du^{2} + \eta_{\mu\nu}d\tilde{x}^{\mu}d\tilde{x}^{\nu}\right]\label{5dmetric}
\end{align}
where $\zeta(u)$ is the inverse scale factor, related to the Einstein-frame warp factor by $\zeta(u)=e^{-A(u)}$ \cite{Ballon-Bayona:2017sxa}. 
To simplify the notation, we introduce the dimensionless coordinates $u$ and $\tilde{x}$, which are related to the usual coordinates $z$ and $x$ by $u=\Lambda z$ and $\tilde{x}=\Lambda x$. The Einstein-dilaton field equations evaluated at the metric \eqref{5dmetric} reduce to
\begin{align}
    \zeta^{\prime\prime} - \dfrac{4}{9}\zeta \Phi^{\prime\,2} = 0\,,\label{edeq1}\\
\dfrac{8}{3}\zeta^{2}\left[\Phi^{\prime\prime} - 3\dfrac{\zeta^{\prime}}{\zeta}\Phi^{\prime}\right] + \dfrac{dV}{d\Phi} = 0\,,\label{edeq2}\\
V - \zeta^{-5}(\zeta^{-3})^{\prime\prime} = 0\,.\label{edeq3}
\end{align}
where $' = d/du$ and $\Phi =\Phi(u)$. Equation \eqref{edeq1} allows us to either prescribe the dilaton field to determine the corresponding inverse scale factor, or specify the inverse scale factor to obtain the associated dilaton. Given this, we employ two analytical Einstein-dilaton models to investigate the electromagnetic form factors of vector mesons. For the first model, we consider a dilaton field that is quadratic in the radial coordinate, motivated by the soft wall model \cite{Karch:2006pv} and obtain the corresponding inverse scale factor, see also \cite{Li:2013oda}. For the second model, we consider a quadratic deformation of the warp factor, motivated by \cite{Gursoy:2007er}, and determine the associated dilaton field. Thus, the Einstein-dilaton models that we will use in this work are given by
\begin{align}
    \Phi_{I}(u) =  u^{2} &, \quad \zeta_{I}(u) =\Gamma\left(\dfrac{5}{4}\right)3^{1/4}\sqrt{u}\,I_{1/4}\left(\dfrac{2}{3}u^{2}\right)\,,\label{modelI} \\
   \Phi_{II}(u) = \dfrac{1}{2}u\sqrt{9 + 4  u^{2}} + \dfrac{9}{4}\sinh^{-1}\left(\dfrac{2}{3}u\right) &, \quad \zeta_{II}(u) =u\exp\left(\dfrac{2}{3} u^{2}\right)\,,\label{modelII}
\end{align}
where the dilaton and inverse scale factor in \eqref{modelI} and \eqref{modelII} are referred to as Einstein-dilaton models I and II, respectively. Near the boundary (UV regime) the dilaton vanishes and the scale factor must be such that the metric is asymptotically $AdS_{5}$, leading to
\begin{align}
    \Phi(u\to 0) = 0,\qquad \zeta(u\to 0) \sim u\,.\label{adslimit}
\end{align}

\begin{figure}[htp!]
    \centering
    \begin{subfigure}{0.45\textwidth}
        \centering
        \includegraphics[width=\textwidth]{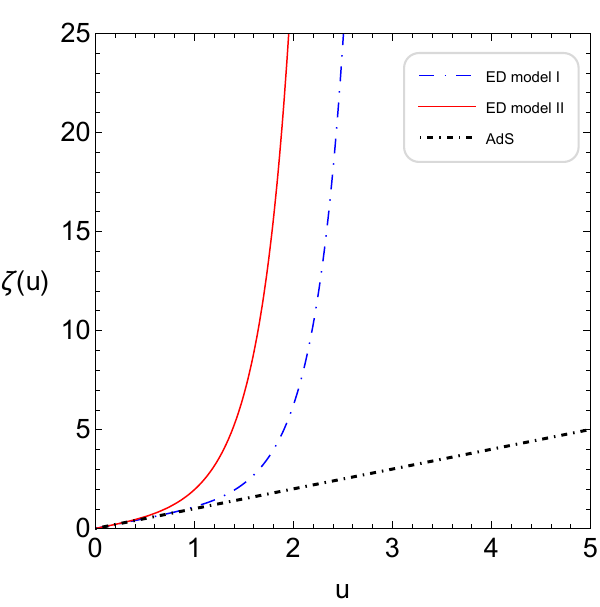}
    \end{subfigure}
    \hspace{0.05\textwidth}
    \begin{subfigure}{0.45\textwidth}
        \centering
        \includegraphics[width=\textwidth]{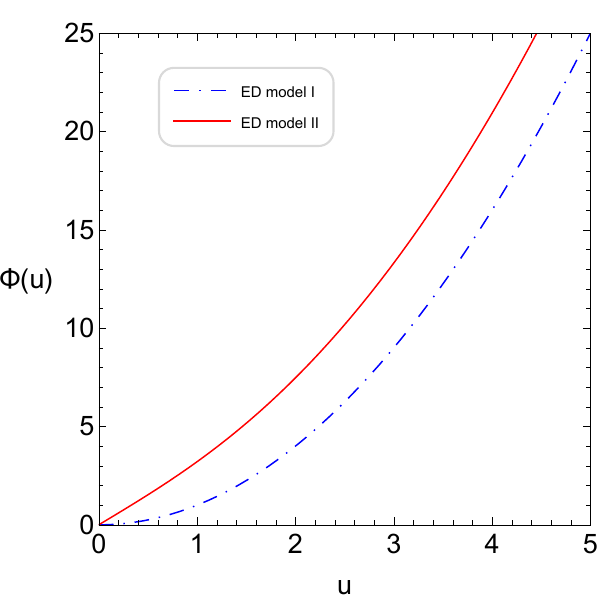}
    \end{subfigure}
    \caption{{\bf Left panel}: Inverse scale factor in the Einstein-frame as a function of $u$ dimensionless coordinate for Einstein dilaton models I (blue dot-dashed line) and  II (red solid line). We also plot the AdS limit (black dotted line), in which $\zeta(u) = u$. {\bf Right panel}: Dilaton field as function of the dimensionless coordinate $u$ for Einstein dilaton models I (blue dot-dashed line) and II (red solid line).}
    \label{Plot:Scalefactordilaton}
\end{figure}

 The left panel of Fig. \ref{Plot:Scalefactordilaton} shows the behavior of inverse scale factor $\zeta(u)$ for Einstein-dilaton models I, II, and the AdS limit according to \eqref{adslimit}.  The right panel of Fig. \ref{Plot:Scalefactordilaton} displays the dilaton field for models I and II.

\begin{figure}[htp!]%
    \centering
    {{\includegraphics[width=7cm]{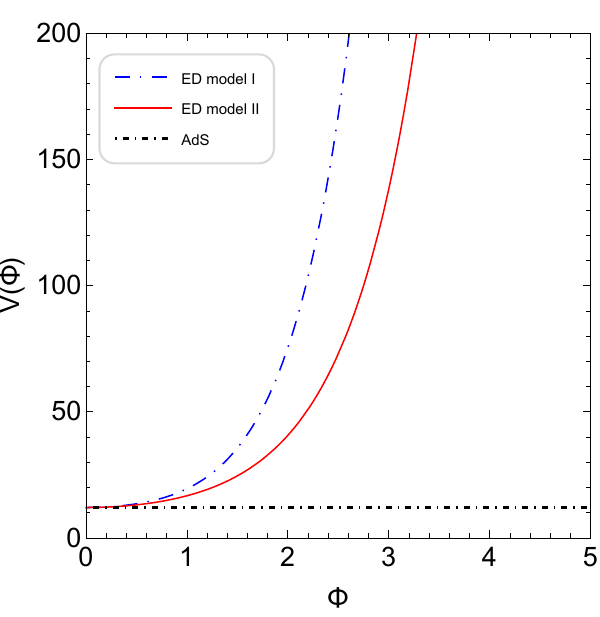} }}%
    
    \caption{The dilaton potential $V(\Phi)$ in the Einstein-frame for models I (blue dot-dashed line) and II (red solid line). We also plot the AdS limit (black dotted line), in which the dilaton potential is constant, $V = 12$, when the dilaton field goes to zero, $\Phi\to 0$.}
    \label{Plot:dilatonpotential}
\end{figure}

From the inverse scale factor and dilaton specified in \eqref{modelI} and \eqref{modelII}, one can reconstruct the potential of the models using \eqref{edeq3}. Figure \ref{Plot:dilatonpotential} exhibits the potentials for models I, II and AdS limit. The latter reduces to $V = 12$, which is associated with the $5d$ negative cosmological constant.

The Einstein-dilaton holographic QCD approach that we follow is compatible with the confinement criterion based on the Wilson loop area law, as discussed in detail in Ref.~\cite{Gursoy:2007er}. In the context of AdS/CFT, Wilson loops are related to minimum surfaces of the Nambu–Goto action \cite{Maldacena:1998im, Kinar:1998vq}, which is defined in the string frame. In Einstein-dilaton holographic QCD, the warp factor in the string frame is given by
\begin{align}
    A_{s}(u) = A(u) + \dfrac{2}{3}\Phi(u)\,. \label{sfwarp}
\end{align}
Considering a rectangular Wilson loop one can describe the potential energy of a heavy static quark-antiquark pair and confinement is satisfied if  the potential grows linearly for long distances. Following \cite{Gursoy:2007er}, one can apply the confinement criterion \cite{Kinar:1998vq} in Einstein-dilaton holographic QCD backgrounds and find that confinement is satisfied as long as the function $f(u) = \exp(2A_{s})$ exhibits a non-zero minimum. The Einstein-dilaton models I and II considered in this paper satisfy this confinement criterion, as shown explicitly in \cite{Ballon-Bayona:2024yuz}.  

In the following subsection, we review the action for the vector meson sector in Einstein-dilaton holographic QCD and the results for the spectrum and decay constants obtained in  \cite{Ballon-Bayona:2024yuz}. Then in the next section we will focus  on the cubic interaction terms and present a systematic study of the vector meson electromagnetic form factors. 

\subsection{Vector mesons}\label{subsec:vectormesons}

\subsubsection{The action and field equations}

We are interested in describing vector mesons made of light quarks $u$ and $d$.  We will assume isospin symmetry ($m_u=m_d$) and therefore the corresponding vector isospin current $J^{\mu, a}(\tilde{x}) =\bar{q}(\tilde{x})\gamma^{\mu}\tau^{a}q(\tilde{x})$ is conserved, where $\tau^{a}$ are the generators of the isospin group $SU(2)$. In holographic QCD the vector isospin current is mapped to a five dimensional non-Abelian gauge field \cite{Erlich:2005qh,Grigoryan:2007vg,Grigoryan:2007my}. The corresponding Yang-Mills action coupled minimally to the metric and the dilaton can be written in the string frame as \footnote{For higher order corrections to the 5d Yang-Mills action in holographic QCD see for instance Ref.\cite{Grigoryan:2007iy}.}
\begin{equation}
S = - \int d^4 x dz \sqrt{-g_{s}} \, e^{-\Phi}\,
{\rm Tr} \Big  ( \frac{1}{2 g_5^2}
V_{mn}^2 \Big ) \, ,\label{vecaction1}
\end{equation}
where 
\begin{align}
V_{mn} \equiv v_{mn}  - i [V_m , V_n]  
\quad , \quad
v_{mn} \equiv \partial_m V_n - \partial_n V_m \,,\label{Eq:FieldStrength}
\end{align}
being $V_m = V_m^a t^a$ the non-Abelian gauge field. Note that $v_{mn}$ can be written as $v_{mn}^a t^a$ where $v_{mn}^a = \partial_m V_n^a - \partial_n V_m^a$ is the Abelian field strength.

In the 5d Yang-Mills action \eqref{vecaction1}, the  coupling is fixed as  $g^{2}_{5} = 12\pi^{2}/N_{c}$ in order to reproduce the current correlator of large $N_c$ perturbative QCD in the small-distance regime~\cite{Erlich:2005qh}. The 5d metric $g^{s}_{mn}$ is defined in the string frame and can be written as 
\begin{align}
    g_{mn}^{s} = e^{2A_{s}}\eta_{\hat{m}\hat{n}}\,,\label{metricstring}
\end{align}
where $A_s$ is the string frame warp factor given in \eqref{sfwarp} and $\eta_{\hat{m}\hat{n}}$ is the 5d Minkowski metric and we use hatted  indices to distinguish the Minkowski space metric from the curved space one. The 5d Yang-Mills action can be decomposed as 
\begin{align}
    S = S_2 + S_3 + S_4 \, , 
\end{align}
where
\begin{align}
S_2 &= - \int d^5 x \sqrt{-g_{s}} \,    e^{-\Phi} \left( \frac{1}{4 g_5^2} {v_{mn}^a}^2  \right) \nonumber \\
&= - \int d^4 x \int dz \,   e^{A_s - \Phi} \,
\left( \frac{1}{4 g_5^2} {v_{\hat m \hat n}^a}^2   \right) \,  , \label{S2}
\end{align}
\begin{align}
S_3 &=  - \int d^5 x \sqrt{ - g_{s}} \,    e^{-\Phi}\left(  \frac{1}{2g_5^2} f^{abc}  v^{mn}_a\,V_m^b V_n^c \right) \nonumber \\
&=  - \int d^4 x  \int dz \,   e^{A_s - \Phi} \, \left( \frac{1}{2g_5^2} f^{abc}  v^{\hat m \hat n}_a V_{\hat m}^b V_{\hat n}^c  
 \right) \,,  \label{S3}
\end{align}
\begin{align}
S_4 &=  - \int d^5 x \sqrt{ - g_{s}} \,    e^{-\Phi} \left(  \frac{1}{4g_5^2} f^{abe} f^{cde} V^m_a V^n_b V_m^c V_n^d \right) \nonumber \\
&=  - \int d^4 x  \int dz \,   e^{A_s - \Phi} \, \left(  \frac{1}{4g_5^2} f^{abe} f^{cde} V^{\hat m}_a V^{\hat n}_b V_{\hat m}^c V_{\hat n}^d \right) \,,  \label{S4}
\end{align}
and $f^{abc}$ are the structure constants. The non-hatted indices are raised using the original metric while the hatted indices are raised using the flat space Minkowski metric. From the action \eqref{S2} we will obtain the spectrum of vector mesons while the terms \eqref{S3} and \eqref{S4} give rise to three and four point current correlators, respectively.

\subsubsection{Spectrum and decay constants}
Here we follow closely the derivations given in Ref. \cite{Ballon-Bayona:2024yuz}. Varying the action \eqref{S2} we obtain the field equations 
\begin{align}
     \partial_{\hat{m}}\left(e^{A_{s} - \Phi}v^{\hat{m}\hat{n}\,a}\right) = 0\,.\label{veceqmotionabelian}
\end{align}
The field equations are  invariant under the following gauge symmetry
\begin{align}
    V_{\hat{m}}^{a}\rightarrow  V_{\hat{m}}^{a} - \partial_{\hat{m}}\lambda_{V}^{a}\,.\label{gaugesymmetry}
\end{align}

The vector field and the derivatives can be decomposed into their components as $V_{\hat{m}}^{a} = (V_{u}^{a}, V_{\hat{\mu}}^{a})$ and $\partial_{\hat{m}} = (\partial_{u}, \partial_{\hat{\mu}})$. 
The $4d$ vector gauge field $V_{\hat{\mu}}$ can further be decomposed into irreducible representations of the Lorentz group, $SO(1,3)$, as $
    V_{\hat{\mu}}^{a} = V_{\hat{\mu}}^{\perp,a} + \partial_{\hat{\mu}}\xi^{a}$, 
where $V_{\hat{\mu}}^{\perp,a}$ denotes the transverse vector field, which satisfies the condition $\partial_{\hat{\mu}}V^{\hat{\mu}, a}_{\perp} = 0$ and describes the $\rho$ meson family, while $\xi^{a}$ is unphysical  mode that can be set to zero. Using  the gauge transformation \eqref{gaugesymmetry} to fix $V_{u}^{a} = 0$, the vector field equations \eqref{veceqmotionabelian} reduce to 
\begin{align}
    \left[\partial_{u} + A_{s}^{\prime} - \Phi^{\prime}\right]\partial_{u}V^{\hat{\mu}, a}_{\perp}  + \Box V^{\hat{\mu}, a}_{\perp} = 0\,.\label{transversevectorfield}
\end{align}
We will assume from now on that $V_{a}^{\hat{\mu}} \equiv V_{a}^{\hat{\mu}, \perp}$. From the action \eqref{S2}, we extract the on-shell ($on$) contribution, which can be expressed as 
\begin{align}
     S_{2}^{on} &=  \int d^4 x \Big [ \frac{1}{2 g_5^2} e^{A_s - \Phi} (\partial_u V^{\hat \mu}_a) V_{\hat \mu}^a \Big ]_{u= \epsilon}\,.\label{vecaction1onshell}
\end{align}
The vector gauge field in real space can be written as
\begin{align}
    V_{\hat{\mu}}^{a}(u,\tilde{x}) = \int d^{4}\tilde{y}\, K_{\hat{\mu}\hat{\nu}}^{ab}(u, \tilde{x}; \tilde{y}) V_{b}^{\hat{\nu},(0)}(\tilde{y})\,.\label{realbdypropagator}
\end{align}
where $K_{\hat{\mu}\hat{\nu}}^{ab}(u, \tilde{x}; \tilde{y})$ is the bulk to boundary propagator. Substituting \eqref{realbdypropagator} into the on-shell action \eqref{vecaction1onshell}, we obtain the 2-point current correlator in 4d given by
\begin{align}
    G_{\hat \mu \hat \nu}^{cd} (\tilde{x} - \tilde{y}) &= \langle J_{\hat \mu, c} (\tilde{x}) J_{\hat \nu,d} (\tilde{y})  \rangle = \frac{ \delta^2 S_V^{on}}{ \delta V^{\hat \mu,0}_c(\tilde{x}) \delta V^{\hat \nu,0}_d(\tilde{y})} = \frac{1}{g_5^2} \Big [ e^{A_s - \Phi} \partial_u  K_{\hat \mu \hat \nu}^{cd} (u,\tilde{x};\tilde{y}) \Big ]_{u= \epsilon}\,.\label{dictionvector}
\end{align}

\begin{figure}[ht!]
    \centering
    \begin{subfigure}{0.45\textwidth}
        \centering
        \includegraphics[width=\textwidth]{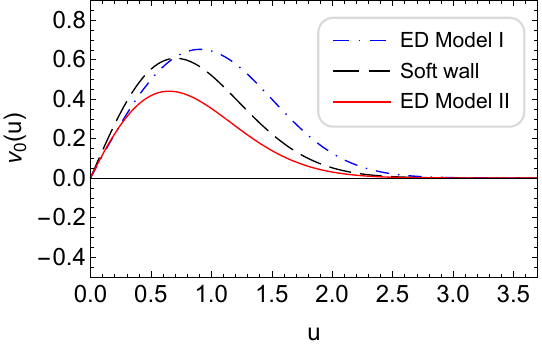}
    \end{subfigure}
    \hspace{0.05\textwidth}
    \begin{subfigure}{0.45\textwidth}
        \centering
        \includegraphics[width=\textwidth]{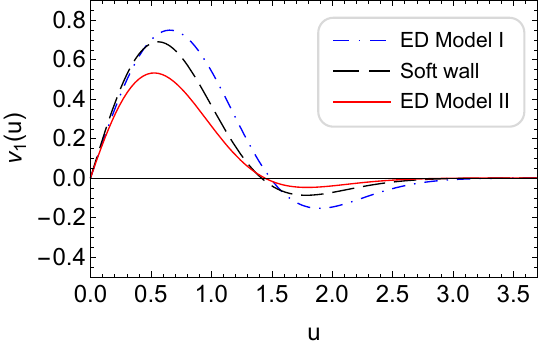}
    \end{subfigure}

    \vspace{0.5cm}

    \begin{subfigure}{0.45\textwidth}
        \centering
        \includegraphics[width=\textwidth]{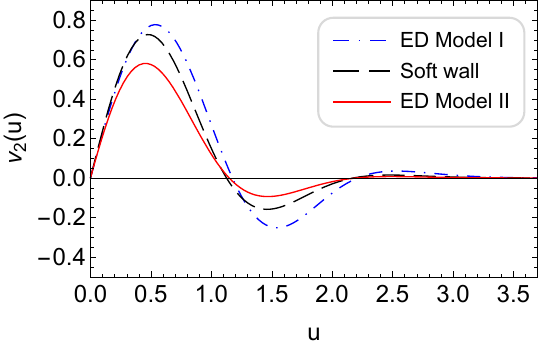}
    \end{subfigure}
    \hspace{0.05\textwidth}
    \begin{subfigure}{0.45\textwidth}
        \centering
        \includegraphics[width=\textwidth]{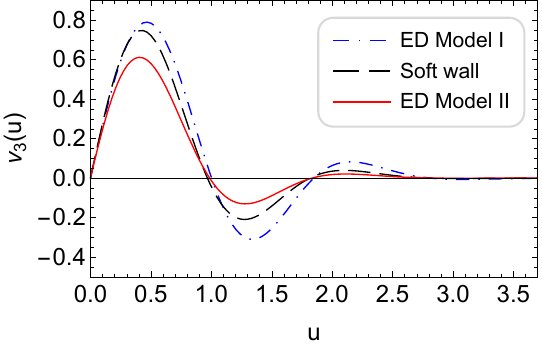} 
    \end{subfigure}

    \caption{Normalizable modes (or Sturm-Liouville modes) as a function of $u$ dimensionless coordinate for models I (blue dot-dashed line), II (red solid line), compared with the Soft wall model (black dashed line): (i) Mode $v_{0}$ (upper left panel), 
    (ii) Mode $v_{1}$ (upper right panel), 
    (iii) Mode $v_{2}$ (lower left panel), and 
    (iv) Mode $v_{3}$ (lower right panel).}
    \label{Plot:vnmodes}
\end{figure}

Performing a 4d Fourier transform, we can write the bulk to boundary propagator as
\begin{align}
    K_{\hat{\mu}\hat{\nu}}^{ab}(q^{2}, u) = P^{\hat \mu \hat \nu} (q) \delta^{ab} \, V( q^{2}, u) \, , \label{bulktoboundarytransverse}
\end{align}
where   
\begin{equation}
P^{\hat \mu \hat \nu} (q) \equiv \eta^{\hat \mu \hat \nu} - \dfrac{q^{\hat \mu}q^{\hat \nu}}{q^{2}}  \, , \label{projector}  
\end{equation}
is the transverse projector and  $V(q^2,u)$ is the scalar piece of the bulk to boundary propagator, which satisfies the equation
\begin{align}
    \left[\left(\partial_{u} + A_{s}^{\prime} - \Phi^{\prime}\right)\partial_{u} - q^{2}\right]V(q^{2}, u) = 0\,.\label{bulk to boundary propagator}
\end{align}
The  Green's function for the differential equation in \eqref{bulk to boundary propagator} and its relation to the bulk to boundary propagator can be found using Sturm - Liouville theory. The result takes the following form 
\begin{align}
    G(q^{2}, u;u^{\prime}) = - \sum_{n} \dfrac{v^{n}(u)v^{n}(u^{\prime})}{q^{2} + m_{v^{n}}^{2}}\,,\quad V(q^{2}, u^{\prime}) = -\left[e^{A_{s} - \Phi}\partial_{u} G(q^{2},u;u^{\prime})\right]_{u = \epsilon}\label{vecgreenfunction} 
\end{align}
where the Sturm - Liouville modes $v^{n}$ satisfy the orthonormality condition 
\begin{align}
    \int du\,e^{A_{s} - \Phi}v^{n}(u)v^{m}(u) = \delta^{mn}\,,
\end{align}
and the eigenvalue equation
\begin{align}
    \left[\partial_{u}\left(e^{A_{s} - \Phi}\partial_{u}\right) + m_{v^{n}}^{2}e^{A_{s} - \Phi}\right]v^{n}(u) = 0\label{modeseigenvalueequation}\,.
\end{align}
The asymptotic behavior for small $u$ of the normalizable modes $v^{n}(u)$ is given by
\begin{align}
    v^{n}(u) = N_{v^{n}}u^{2} + \dots\label{Eq:AsympBWF}
\end{align}
where $N_{v^{n}}$ are the normalization constants. The Schrödinger potential and the eigenfunctions of $\rho$ mesons were analyzed in detail in Ref.~\cite{Ballon-Bayona:2024yuz}. Figure~\ref{Plot:vnmodes} shows the behavior of the normalizable modes for model I (blue dot-dashed line), model II (red solid line), and the soft wall model (black dashed line). The four panels correspond to the modes $v_0$ (upper left), $v_1$ (upper right), $v_2$ (lower left), and $v_3$ (lower right).

The results for the spectrum of vector mesons in Einstein dilaton I and II models, compared with experimental data, the  hard-wall and the soft-wall models are presented in table \ref{Table:VectorMesonMasses}. As explained in Ref.~\cite{Ballon-Bayona:2024yuz},  we consider the ratios between the masses of the excited states and the ground state of the vector mesons, $m_{\rho^{n}}/m_{\rho^{0}}$, so that it is not necessary to fix the mass scale parameter $\Lambda$. 

\begin{table}[ht]
\centering
\begin{tabular}{l |c|c|c|c|c}
\hline 
\hline
Ratio & Model I & Model II & Soft wall  & Hard wall & Experimental \\
\hline 
 $m_{\rho^1}/m_{\rho^0}$ & 1.591  & 1.340   & 1.414  & 2.295 &  $1.652 \pm 0.048$  \\
 $m_{\rho^2}/m_{\rho^0}$ & 2.015  & 1.611  & 1.732  & 3.598   & $1.888 \pm 0.032$  \\
 $m_{\rho^3}/m_{\rho^0}$ & 2.365  &1.843  & 2 & 4.903  & $2.216\pm 0.026$ \\
 $m_{\rho^4}/m_{\rho^0}$ & 2.670  & 2.049 &  2.236 & 6.209   & $2.443 \pm 0.072$  \\
 $m_{\rho^5}/m_{\rho^0}$ & 2.944  &  2.236 &  2.450 & 7.514   & $2.727 \pm 0.265$  \\
\hline\hline
\end{tabular}
\caption{
Ratio of vector meson masses $m_{\rho^n}/m_{\rho^0}$ for the first excited states $n=1,..,5$ in the Einstein-dilaton models I and II, the soft wall model and the hard wall model, compared against experimental results.  The experimental result for $m_{\rho^1}$ was taken from \cite{OBELIX:1997zla} and the experimental results for the other states were obtained from PDG \cite{Workman:2022ynf}, including the mass of the fundamental state $m_{\rho^0} = 0.776 \pm 0.001 \, {\rm GeV}$. The numerical error in our computations of mass ratios in Einstein-dilaton models I and II was of the order of $10^{-6}$.
}
\label{Table:VectorMesonMasses}
\end{table}
Using \eqref{dictionvector}, \eqref{bulktoboundarytransverse} and \eqref{vecgreenfunction}, we obtain the 2-point current correlator in momentum space 
\begin{align}
     G_{ \hat \mu \hat \nu}^{ab}(q^{2}) = P_{\hat \mu \hat \nu} (q) \delta^{ab} \sum_n \frac{ F_{v^n}^2 }{q^2 + m_{v^n}^2}\quad  \,,\quad F_{v^n}= \frac{1}{g_5} \Big [ e^{A_s - \Phi} \partial_u  v_n(u)  \Big ]_{u= \epsilon} =\dfrac{2}{g_{5}}N_{v^{n}} \, , \label{2pointvecdecay}
\end{align}
where $P_{\hat \mu \hat \nu} (q)$ is the transverse projector and $F_{v^n}$ are the decay constants of vector mesons, defined through the matrix element relation
\begin{align}
    \bra{0}J^{\hat \mu,\,a}(0)\ket{v^{n,\,b}(p,\epsilon)} = \delta^{ab}\,F_{v^{n}}\,\epsilon^{\hat \mu}(p,\epsilon) \, ,\label{decayconstant}
\end{align}
which corresponds to the decay of the $\rho$ meson family to the hadronic vacuum. In \eqref{decayconstant}  $\epsilon^{\hat \mu}(p,\epsilon)$ is the polarization vector. The results for the decay constants of vector mesons are presented in table~\ref{Table:VectorMesonDecayConstants}, where we compare the predictions of  Einstein–dilaton models I and II with experimental data, the hard wall model and the soft-wall model \footnote{In the hard-wall model Neumann boundary condition were imposed at the wall.}.

\begin{table}[ht]
\centering
\begin{tabular}{l |c|c|c|c|c}
\hline 
\hline
Ratio & Model I & Model II & Soft wall  & Hard wall &  Experimental \\
\hline 
$\sqrt{F_{\rho^0}}/m_{\rho^0}$  &  0.3719 & 0.283  & 0.3355  & 0.4246 &  $0.446 \pm 0.0019$  \\
$\sqrt{F_{\rho^1}}/m_{\rho^0}$ & 0.4704  & 0.3407 & 0.3989  & 0.7946    & -  \\
$\sqrt{F_{\rho^2}}/m_{\rho^0}$ & 0.5298  & 0.3798  & 0.4415  & 1.114    & - \\
$\sqrt{F_{\rho^3}}/m_{\rho^0}$ & 0.5741  & 0.41   & 0.4744  & 1.405    & - \\
$\sqrt{F_{\rho^4}}/m_{\rho^0}$ & 0.61  & 0.4351  & 0.5017  & 1.677     & - \\
\hline\hline
\end{tabular}
\caption{
Dimensionless ratios $\sqrt{F_{\rho^n}}/m_{\rho^0}$  for vector meson decay constants in the Einstein-dilaton models I and II, the soft wall model and the hard wall model, compared against the experimental result. The experimental result was obtained using $\sqrt{F_{\rho^0}} = 0.3462 \pm 0.0014 \, {\rm GeV}$  \cite{Donoghue:1992dd} and $m_{\rho^0} = 0.776 \pm 0.001 \, {\rm GeV}$ \cite{Workman:2022ynf}. The numerical error in our computations of $\sqrt{F_{\rho^n}}/m_{\rho^0}$ in Einstein-dilaton models I and II was of the order of $10^{-3}$.
}
\label{Table:VectorMesonDecayConstants}
\end{table}

The electromagnetic form factor of vector mesons requires specifying the values of the  infrared mass scale parameter $\Lambda$, associated with conformal symmetry breaking. Table \ref{Table:kparameter} provides the values of the parameter $\Lambda$ fixed using the ground state mass of the $\rho$ meson  in Einstein-dilaton models I and II, the soft-wall model and the hard-wall model. The values presented in this table will be used in the next section where we calculate the electromagnetic form factors.

\begin{table}[htp!]
\centering
\begin{tabular}{c |c|c|c|c}
\hline 
\hline
 Parameter & Model I & Model II & Soft wall  & Hard wall \\
\hline 
 $\Lambda$ & $0.481\, \text{GeV}$  &   $0.348\, \text{GeV}$  &  $0.388\, \text{GeV}$  &  $0.323\, \text{GeV}$ \\
\hline\hline
\end{tabular}
\caption{Values of the infrared mass scale parameter $\Lambda$ for each holographic model, determined from the experimental value of the ground-state $\rho$ meson mass.}
\label{Table:kparameter}
\end{table}

\section{Vector meson electromagnetic form factors in Einstein-dilaton holographic QCD}\label{Sec:FormFactors}

In this section, we will obtain the three-point correlation function of isospin currents and present a systematic description of the electromagnetic form factors in Einsten-dilaton holographic QCD, focusing on the models I and II introduced in section \ref{Sec:VMHQCD}. In subsection \ref{threepointfunctionvec} we compute the three-point correlator in Einstein-dilaton holographic QCD using the bulk to boundary propagators. Performing a spectral decomposition we find a holographic dictionary for the electromagnetic form factors valid for the elastic and non-elastic case. In  subsection \ref{veceletromagneticformfactors} we evaluate the elastic and transition form factors and provide a full description of the strong couplings of vector mesons. In subsection \ref{subsec:currentmatrixelem} we present our holographic dictionary for the current matrix element obtained from the spectral decomposition of the three-point current correlator. This result is confirmed by an independent method based on the Kaluza–Klein expansion presented in Appendix \ref{App:kkexpansion}. In subsection \ref{electricmagneticquadrupole} we compute the (elastic) electric, magnetic, and quadrupole form factors and radii. We also compute the position of the zero associated with the charge form factor, as well as the magnetic and quadrupole moments, and compare our results with those obtained from Lattice QCD and other approaches, such as hard-wall, soft-wall, and Dyson–Schwinger models. 

\subsection{Three point function of isospin currents}\label{threepointfunctionvec}

\begin{figure}[htp!]%
    \centering
    {{\includegraphics[width=5cm]{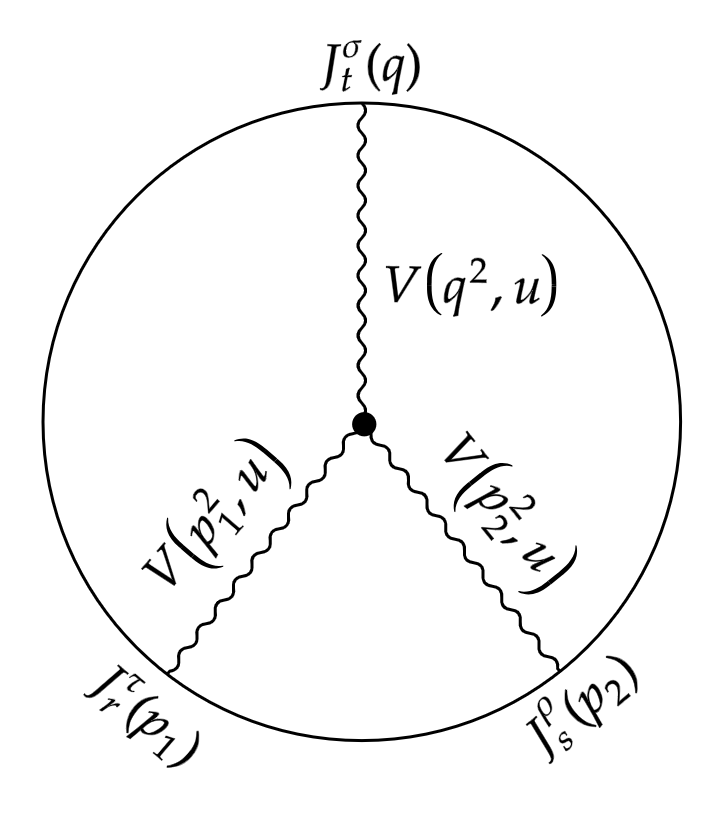}}}%
    \caption{Witten diagram for the 3-point current correlator. The edge represents the conformal boundary of $AdS_{5}$ spacetime,  while its interior represents the bulk region of the asymptotically $AdS_{5}$ spacetime, where the 3-vertex interaction between the bulk to boundary propagators $V(p_{1}^2,u)$, $V(p_{2}^2,u)$ and $V(q^2,u)$ takes place.}
    \label{Plot:WittenVM}%
\end{figure}

In order to compute the three-point function of isospin currents, it is useful to rewrite the cubic interaction  \eqref{S3} in the following form 
\begin{align}
    S_3 = -   \int d^{4}\tilde{x}\, du \, e^{A_{s} - \Phi} \, \dfrac{1}{2g_{5}^{2}} \,f_{abc} \, \Big [ v_{a}^{\hat{\mu}\hat{\nu}}V_{\hat{\mu}}^{b}V_{\hat{\nu}}^{c} + 2 v_{u\hat{\mu}}^{a}V_{u}^{b}V_{c}^{\hat{\mu}} \Big ] \,.\label{vecaction2}
\end{align}
 Using the gauge $V_{u}^{a} = 0$, the second term in the action vanishes and does not contribute to the three-point correlator. 
As shown in the previous section, the longitudinal part of $V_{a}^{\hat{\mu}}$ can be set to zero. We will assume from now on that $V_{a}^{\hat{\mu}} \equiv V_{a}^{\hat{\mu}, \perp}$. Then the action \eqref{vecaction2} becomes
\begin{align}
S_{V V V} &= - \int d^4 x \int du \, e^{A_s - \Phi} \, \frac{1}{g_5^2} f_{abc} \partial^{\hat \mu} V^{a, \hat \nu} V_{ b, \hat \mu} V_{c, \hat \nu}  \,, 
\end{align}
which in momentum space takes the following form 
\begin{align}
S_{V V V} &= - \int d^4 q' \int d^4 p'_1 \int d^4 p'_2 \,  (2 \pi)^4 \, \delta^4 (p'_2 + p'_1 + q')  \int du \, e^{A_s - \Phi} \nonumber \\
&\times \frac{1}{g_5^2} f_{abc} ( - i q'^{\hat \mu} )  V^{a , \hat \nu} (q',u)   V_{b , \hat \mu} (p'_1,u)  V_{c , \hat \nu} (p'_2,u)  \,. \label{SVVVmom}
\end{align}
The vector gauge field \eqref{realbdypropagator} in momenta space can be written as
\begin{align}
    V_{\hat \mu}^{a}(p,u) = K_{\hat \mu \hat \nu}^{ab}(p^2,u)V^{(0),\hat \nu}_{b}(p)\label{gauge5d}
\end{align}
Plugging \eqref{gauge5d} and \eqref{bulktoboundarytransverse} into \eqref{SVVVmom} and varying the action with respect to the respective sources, we obtain the 3-point current correlator 
\begin{align}
     \langle J^{ \hat \tau}_{r}(p_{1})J^{\hat \sigma}_{t}(q)J^{\hat \rho}_{s}(-p_{2})\rangle &= if_{rst} P^{\hat \tau  \hat \tau'}(p_{1}) P^{\hat \rho  \hat \rho'}(-p_{2}) P^{\hat \sigma  \hat \sigma'}(q) \Big [\eta_{\hat \tau' \hat \rho'} \left(p_1 + p_2  \right)_{\hat \sigma'}  \nonumber \\
     &+ 2 \left(\eta_{\hat \tau' \hat \sigma'}q_{\hat \rho'} - \eta_{\hat \rho' \hat \sigma'}q_{\hat \tau'}\right) \Big ]\mathcal{W}(p_{1}^2,p_{2}^2,q^2) (2\pi)^{4}\delta^{4}(p_{1} + q  - p_{2})\,,\label{3correlatortransverse}
\end{align}
where $P^{\hat \tau \hat \tau'}(p_{1})$, $P^{\hat \rho \hat \rho'}(-p_{2})$ and $P^{\hat \sigma \hat \sigma'}(q)$ are the transverse projectors defined in \eqref{projector}, corresponding to currents $J^{\hat \tau}_{r}(p_{1})$, $J^{\hat \rho}_{s}(-p_{2})$ and $J^{\hat \sigma}_{t}(q)$, respectively. The quantity $\mathcal{W}(p_{1}^2,p_{2}^2,q^2)$, known as the dynamical factor \cite{Grigoryan:2007vg}, is given by
\begin{align}
      \mathcal{W}(p_{1}^2,p_{2}^2,q^2) &= \dfrac{1}{g_{5}^{2}}\int du\, e^{A_{s} - \Phi} \, V(p_{1}^2, u)V(p_{2}^2, u)V(q^2, u)\label{dynamicalfactor}\,.
\end{align}
The presence of transverse projectors in \eqref{3correlatortransverse}, ensures the transversality of the $3$-point correlator. The dynamical factor \eqref{dynamicalfactor} describes the interaction between the scalar pieces of the bulk to boundary propagators $V(p_{1}^2, u)$, $V(p_{2}^2, u)$, and $V(q^2,u)$ dual to the currents $J_{r}^{\hat \tau}(p_{1})$, $J_{s}^{\hat \rho}(p_{2})$ and $J_{t}^{\hat \sigma}(q)$, respectively. This interaction can be represented by a Witten diagram, as shown in Fig. \ref{Plot:WittenVM}.

In the next subsection we will use the dynamical factor in \eqref{dynamicalfactor} to extract the elastic and transition form factors and in subsection \ref{subsec:currentmatrixelem} we will obtain our holographic dictionary for the current matrix element. 

\subsection{Elastic and transition form factors}\label{veceletromagneticformfactors}

The dynamical factor \eqref{dynamicalfactor} contains three bulk to boundary propagators. All of them admit a spectral decomposition, which can be written as \cite{Ballon-Bayona:2024yuz}
\begin{align}
    V(p^{2}, u) = g_{5}\sum_{m = 0}^{\infty}\dfrac{F_{v^m}}{p^{2} + m_{v^{m}}^{2}} \, v^{m}(u) \,.\label{nonnormalizablebulkboundary}
\end{align}
where $v^{m}(u)$ are the the 5d normalizable modes and $F_{v^m}$ the vector meson decay constants. Applying the spectral decomposition \eqref{nonnormalizablebulkboundary} for $V(p_{1}^2, u)$ and $V(p_{2}^2, u)$,  the dynamical factor \eqref{dynamicalfactor} becomes
 \begin{align}
    \mathcal{W}(p_{1}^{2}, p_{2}^{2}, q^{2}) =\sum_{n=0}^{\infty} \sum_{k =0}^{\infty}\dfrac{F_{v^n}F_{v^k}}{(p_{1}^{2} + m_{v^n}^{2})(p_{2}^{2} + m_{v^k}^{2})} \, F_{v^n v^k}(q^{2}) \label{DynFactorDecomp} \, ,
\end{align}
where
\begin{align}
    F_{v^{n}v^{k}}(q^{2}) =\int du\, e^{A_{s} - \Phi}\,V(q^{2}, u)\,v^{n}(u)\,v^{k}(u)\,,\label{vecformfactorgeneral}
\end{align}
is our result for the generalized vector meson form factor which includes the elastic form factor ($k=n$) as well as the transition form factors ($k \neq n$).

\begin{figure}[htp!]%
    \centering
    {{\includegraphics[width=6cm]{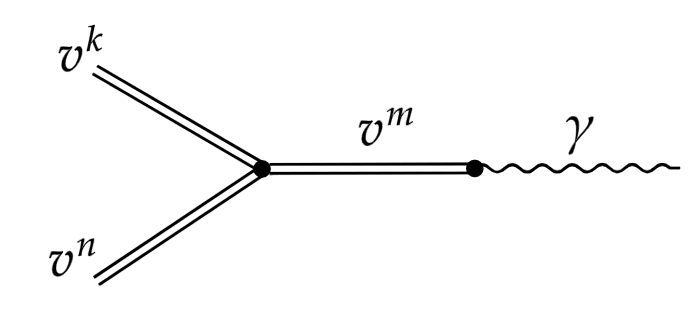}}}%
    \caption{Feynman diagram representing the decomposition of the electromagnetic form factor into the strong couplings between three vector mesons.}%
    \label{Plot:FeynmannVM}%
\end{figure}

Using once more the spectral decomposition \eqref{nonnormalizablebulkboundary}, the generalized form factor \eqref{vecformfactorgeneral} acquires the form
\begin{align}
     F_{v^{n}v^{k}}(q^{2}) = \sum_{m = 0}^{\infty} \dfrac{F_{v^{m}} }{q^{2} + m_{v^{m}}^{2}} \,  g_{v^{m}v^{n}v^{k}} \,.\label{vecnonelasticformfactor}
\end{align}
where
\begin{align}
    g_{v^{m}v^{n}v^{k}} = g_{5}\int du\, e^{A_{s} - \Phi} v^{m}(u)\,v^{n}(u)\, v^{k}(u)\,,\label{couplingvecnk}
\end{align}
represents  the strong couplings between three vector mesons. The result above corresponds to a generalization of vector meson dominance\footnote{Generalization of vector meson dominance in AdS/QCD was originally observed in \cite{Hong:2004sa}.} and is illustrated by the Feynman diagram in Fig. \ref{Plot:FeynmannVM}.

Using  \eqref{couplingvecnk}, we compute the  strong couplings between vector mesons for the hard-wall model, the soft-wall model and the Einstein-dilaton models I and II. The couplings $g_{v^{m}v^{0}v^{0}}$ that correspond to the elastic form factor are displayed in table \ref{table:VectorMesonCouplings}. It is worth noting that the elastic coupling between ground-state vector mesons is the strongest, while the couplings involving higher excited states decrease rapidly for Einstein-dilaton I, II and hard-wall, whereas for soft-wall, the elastic couplings from the second excited state onwards are zero. 
The couplings $g_{v^{m}v^{0}v^{k}}$, with $k=1,2,3$, corresponding to the non-elastic form factor are displayed in table \ref{table:VectorMesonCouplingsnonelastic}. Note that the couplings oscillate due to the transitions between states. Naturally, couplings between closer states are stronger, while couplings between more distant states are weaker. 
It is worth noting that in the soft wall model some meson couplings vanish. This behavior is associated with the orthogonality of the Laguerre polynomials, together with identities involving integrals of two Laguerre polynomials, leading to a selection rule. Specifically, this selection rule allows couplings only between transitions involving vector meson states $v^{n}$ and $v^{k}$ satisfying $|m - k| \le 1$, as described in appendix \ref{vecsoftwall}. The selection rule becomes an approximation when we depart from the soft wall model to Einstein-dilaton models I and II or the hard wall model. 

\begin{table}[htp!]
\centering
\begin{tabular}{c |c|c|c|c|c|c}
\hline 
\hline
Model & Coupling & m=0 & m=1 & m=2 & m=3 & m=4 \\
\hline
Hard-wall & $g_{v^{m}v^{0}v^{0}}$ & $6.863$ & $-1.997$ & $0.0220$ & $-0.0025$ & $0.0005$ \\
\hline
Soft-wall & $g_{v^{m}v^{0}v^{0}}$ & $4\pi \sqrt{2}$ & $- 4\pi$ & $0$ & $0$ & $0$ \\
\hline
Einstein-dilaton I & $g_{v^{m}v^{0}v^{0}}$ & $11.39$ & $-5.923$ & $-0.6428$ & $-0.1645$ & $-0.0519$ \\
\hline
Einstein-dilaton II & $g_{v^{m}v^{0}v^{0}}$ & $31.82$ & $-26.63$ & $2.542$ & $0.5006$ & $0.1804$ \\
\hline\hline
\end{tabular}
\caption{Strong elastic coupling constants between vector mesons for hard-wall, soft-wall, and Einstein-dilaton models I and II.}
\label{table:VectorMesonCouplings}
\end{table}  

\begin{figure}[htp!]%
    \centering
    {{\includegraphics[width=9cm]{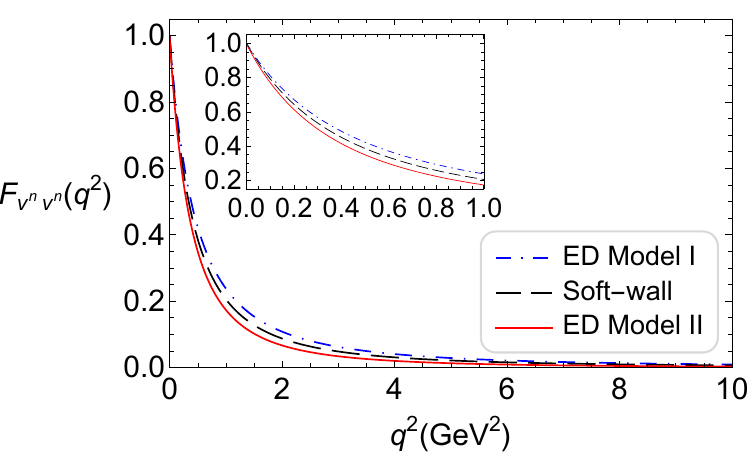} }}%
    
    \caption{Elastic form factors of vector mesons in the Einstein-dilaton models I and II compared against soft-wall model.}
    \label{Plot:Elasticvecformfactors}
\end{figure}

Figure \ref{Plot:Elasticvecformfactors} shows the behavior of the $\rho$-meson elastic form factor for the Einstein-dilaton I and II models, compared with the soft-wall model, as computed from equation \eqref{vecnonelasticformfactor} using the strong elastic couplings listed in table \ref{table:VectorMesonCouplings}. In all three models: Einstein-dilaton I, II, and soft-wall, the form factor is normalized to unity at $q^{2} = 0$. As will be see in the next subsection, this normalization ensures the correct electric charge of the $\rho$ meson. As the momentum transfer increases, these form factors decay to zero, with model II exhibiting the fastest decline, followed by the soft-wall model, while model I displays the slowest decrease.

\begin{table}[htp!]
\centering
\begin{tabular}{c | c | c | c | c | c | c}
\hline\hline
Model & Coupling & m=0 & m=1 & m=2 & m=3 & m=4 \\
\hline

Hard-wall 
& $g_{v^{m}v^{0}v^{1}}$ & -1.997 & 4.700 & -2.148 & 0.03366 & -0.0047 \\
& $g_{v^{m}v^{0}v^{2}}$ & 0.0220 & -2.148 & 4.489 & -2.182 & 0.03777 \\
& $g_{v^{m}v^{0}v^{3}}$ & -0.0025 & 0.0337 & -2.182 & 4.425 & -2.195 \\

\hline

Soft-wall 
& $g_{v^{m}v^{0}v^{1}}$ &  $-4\pi$ & $8\pi\sqrt{2}$ & $-4\pi\sqrt{3}$ & 0 & 0\\
& $g_{v^{m}v^{0}v^{2}}$ & 0 &  $-4\pi\sqrt{3}$ & $12\pi\sqrt{2}$ & $-4\pi \sqrt{6}$ & 0 \\
& $g_{v^{m}v^{0}v^{3}}$ & 0 & 0 &$-4\pi \sqrt{6}$ & $16\pi\sqrt{2}$ & $-4\pi\sqrt{10}$ \\

\hline

Einstein-dilaton I 
& $g_{v^{m}v^{0}v^{1}}$ & - 5.923 & 17.16 & - 8.492 & - 1.103 & - 0.3467 \\
& $g_{v^{m}v^{0}v^{2}}$ & -0.6428 & -8.492 & 22.68 & -10.58 & -1.434 \\
& $g_{v^{m}v^{0}v^{3}}$ & -0.1645 & -1.103 & -10.58 & 27.62 & -12.44 \\

\hline

Einstein-dilaton II 
& $g_{v^{m}v^{0}v^{1}}$ & 31.82 & -26.63 & 2.542 & 0.5006 & 0.1804 \\
& $g_{v^{m}v^{0}v^{2}}$ & 2.542 & -51.26 & 118.4 & -78.02 & 6.698 \\
& $g_{v^{m}v^{0}v^{3}}$ & 0.5006 & 4.567 & -78.02 & 168.2 & -106.6 \\

\hline\hline
\end{tabular}

\caption{Strong non-elastic coupling constants between vector mesons for hard-wall, soft-wall, and Einstein-dilaton models.}

\label{table:VectorMesonCouplingsnonelastic}

\end{table}

\begin{figure}[htp!]
    \centering
    \begin{subfigure}{0.45\textwidth}
        \centering
        \includegraphics[width=\textwidth]{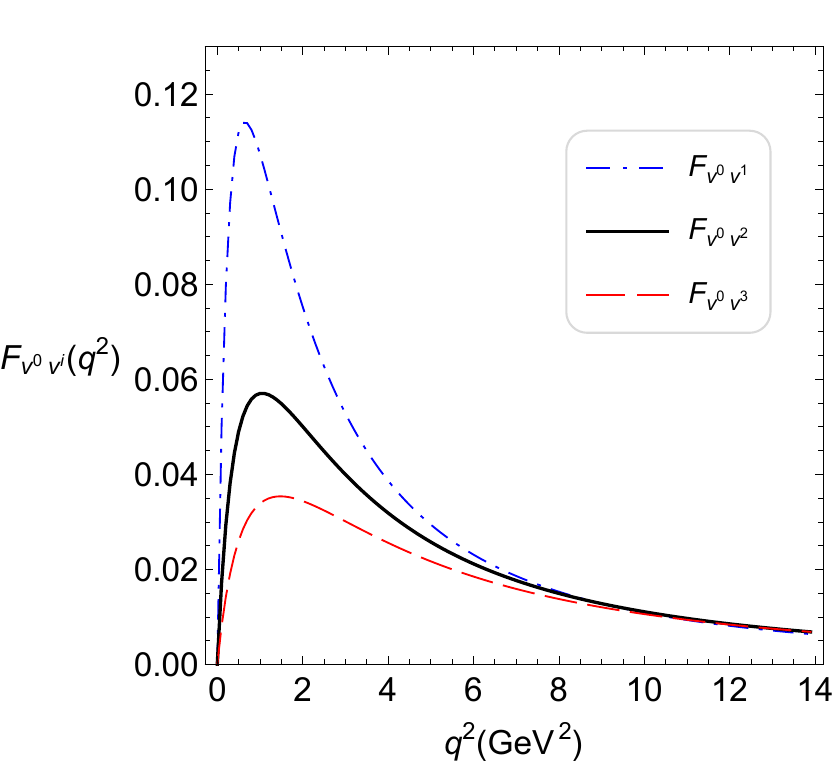}
    \end{subfigure}
    \hspace{0.05\textwidth}
    \begin{subfigure}{0.45\textwidth}
        \centering
        \includegraphics[width=\textwidth]{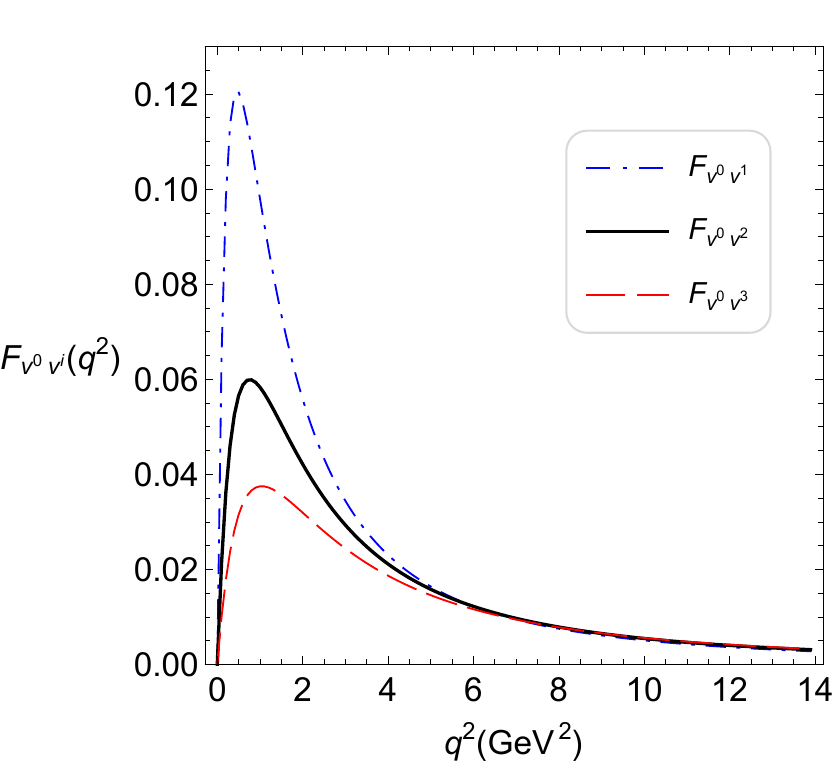}
    \end{subfigure}
    \caption{Transition form factors $F_{v^{0}v^{i}}$ with $i=1,2,3$, of the vector mesons as functions of the squared momentum transfer $q^{2}$. {\bf Left panel}:  Einstein-dilaton I. {\bf Right panel}: Einstein-dilaton II.}
    \label{Plot:NonElasticvecformfactors}
\end{figure}

 Using \eqref{vecnonelasticformfactor} and the non-elastic couplings presented in table \ref{table:VectorMesonCouplingsnonelastic}, we describe the non-elastic form factors of the $\rho$-meson. We consider four states: $v^{0}$, $v^{1}$, $v^{2}$, $v^{3}$, and compute the transition form factors from the fixed initial state $v^0$ to the final states $v^1$, $v^2$, and $v^3$. Figure \ref{Plot:NonElasticvecformfactors} displays the behavior of the transition form factor of vector mesons as a function of the squared momentum transfer $q^{2}$ for Einstein-dilaton models I and II (left 
 and right panels respectively). Note that for both Einstein-dilaton models I and II, the non-elastic form factors are non-monotonic and go to zero in the limit  of large $q^{2}$. For both models at low $q^2$ the form factor $F_{v^0v^1}$ (blue dot-dashed line) provides the leading contribution, followed by $F_{v^0v^2}$ (black solid line) and $F_{v^0v^3}$ (red dashed) decreasing order.
 
\subsection{The current matrix element} \label{subsec:currentmatrixelem}

In the previous subsection, we obtained the three-point function of vector mesons, which contains the dynamical factor. Consider the complete set of intermediate vector-meson states satisfying the completeness relation \cite{Peskin:1995ev}
\begin{align}
    \int \dfrac{d^{3}p}{(2\pi)^{3}}\dfrac{1}{2E_{p}}\sum_{n}\ket{v^{n}(p,\epsilon)}\bra{v^{n}(p,\epsilon)} = 1\,.\label{completeset}
\end{align}
Plugging \eqref{completeset} into the 3-point current correlator (left hand side of \eqref{3correlatortransverse}), we obtain a decomposition of the 3-point current correlator in terms of the current matrix element
\begin{align}
     \langle J^{\tau}_{r}(p_{1})J^{\sigma}_{t}(q)J^{\rho}_{s}(-p_{2})\rangle &= \sum_{n, k}\sum_{\epsilon,\epsilon^{\prime}} \dfrac{F_{v^{n}}F_{v^{k}}}{(p_{1}^{2} + m_{n}^{2})(p_{2}^{2} + m_{k}^{2})}\,f_{rst}\,\epsilon_{\tau}\,{\epsilon'}_{\rho} \nonumber \\
     &\,\bra{v_{r}^{n}(p_{1}, \epsilon)} J^{\sigma}_{t}(0) \ket{v_{s}^{k}(p_{2}^{\prime}, \epsilon^{\prime})}\,(2\pi)^{4}\delta^{4}(p_{1} - p_{2} + q)\,,\label{matrixelement3correlator}
\end{align}
where we used the definition of vector meson decay constants given in \eqref{decayconstant}. Comparing \eqref{matrixelement3correlator} with the right hand side of \eqref{3correlatortransverse}, with the decomposition in \eqref{DynFactorDecomp}, we obtain the current matrix element
\begin{align}
\nonumber \bra{v_{r}^{n}(p_{1}, \epsilon)} J^{\sigma}_{t}(0) \ket{v_{s}^{k}(p_{2}, \epsilon^{\prime})} &= f_{rst} \epsilon_{\tau}\,{\epsilon'}_{\rho}  P^{\tau \tau'}(p_{1})P^{\sigma \sigma'}(q)P^{\rho \rho'}(-p_{2}) \Big [ \eta_{\tau' \rho'} \left(p_1 + p_2 \right)_{\sigma'} \\
&+ 2 \left(\eta_{\tau' \sigma'}q_{\rho'} - \eta_{\rho' \sigma'}q_{\tau'}\right) \Big ]F_{v^n v^k}(q^{2})\,.
\end{align}
The transversality of the vector polarization implies that
\begin{align}
\nonumber   \bra{v_{r}^{n}(p_{1}, \epsilon)} J^{\sigma}_{t}(0) \ket{v_{s}^{k}(p_{2}, \epsilon^{\prime})} &= f_{rst} \epsilon^{\tau'} \epsilon'^{\rho'}  P^{\sigma \sigma'}(q) \Big [ \eta_{\tau'\rho'} \left(p_{1} + p_{2}\right)_{\sigma'}\\
&+ 2 \left(\eta_{\tau' \sigma'}q_{\rho'} - \eta_{\rho' \sigma'}q_{\tau'}\right) \Big ]F_{v^n v^k}(q^{2})\,.\label{matrixelement}
\end{align}

The matrix element \eqref{matrixelement} is a general result obtained from three point function approach and is consistent with the Kaluza-Klein method discussed in the appendix \ref{App:kkexpansion}, see Eq. \eqref{Eq:MatrizElementNElastic}.
\subsection{Electric, magnetic and quadrupole (elastic) form factors}\label{electricmagneticquadrupole}

\begin{figure}[htp]
    \centering
    \begin{subfigure}{0.45\textwidth}
        \centering
        \includegraphics[width=\textwidth]{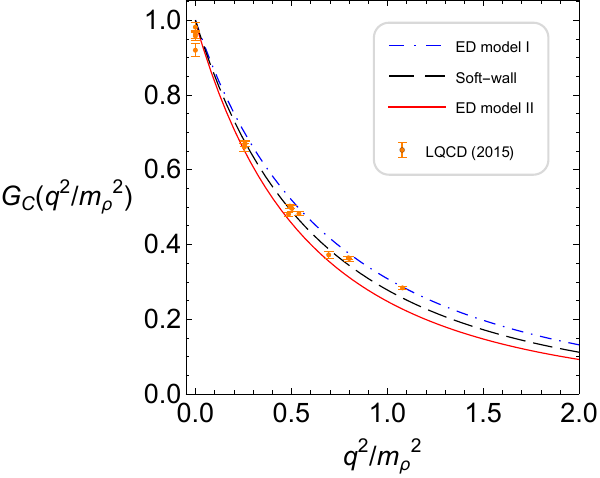}
    \end{subfigure}
    \hspace{0.05\textwidth}
    \begin{subfigure}{0.45\textwidth}
        \centering
        \includegraphics[width=\textwidth]{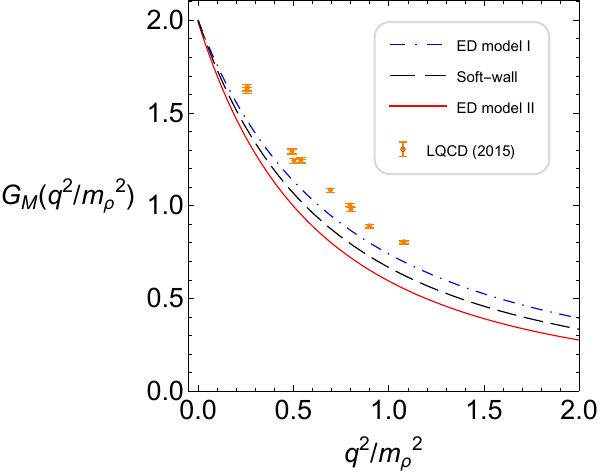}
    \end{subfigure}

    \vspace{0.5cm}

    \begin{subfigure}{0.45\textwidth}
        \centering
        \includegraphics[width=\textwidth]{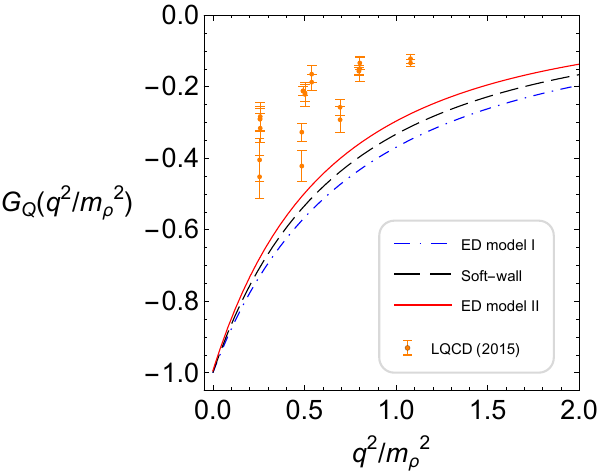}
    \end{subfigure}

    \caption{Electromagnetic form factors of vector mesons for Einstein dilaton I and II models compared with Lattice QCD and soft-wall model: i) Electric form factors (upper left panel) ii) Magnetic form factors (upper right panel) and iii) Quadrupole form factors (lower panel).}
    \label{fig:electromagneticformfactorsvec}
\end{figure}

The matrix element of the electromagnetic current for a massive spin-1 particle, in its most general Lorentz-invariant form \cite{Brodsky:1992px, Haberzettl:2019qpa, Arnold:1979cg, Grigoryan:2007vg}, can be written as
\begin{align}
    \langle v(p_{1}, \epsilon)|J_{em}^{\sigma}(0)|v(p_{2},\epsilon^{\prime})\rangle &= - \epsilon_{\rho}^{\prime} \epsilon_{\tau} \bigg[ 
    \eta^{\rho \tau}\left(p_{1}^{\sigma} - p_{2}^{\sigma}\right)G_{1}(q^{2}) 
    + \left(\eta^{\sigma \rho}q^{\tau} - \eta^{\tau \sigma}q^{\rho}\right) \left(G_{1}(q^{2}) + G_{2}(q^{2})\right) 
    \notag \\ 
    &- \dfrac{1}{m_{v^{m}}^{2}}q^{\rho}q^{\tau}\left(p_{1}^{\sigma} + p_{2}^{\sigma}\right)G_{3}(q^{2}) 
    \bigg]\,,
    \label{matrixelementVM}
\end{align}
where $G_{1}(q^{2})$,  $G_{2}(q^{2})$ and  $G_{3}(q^{2})$ are form factors. Note that both \eqref{matrixelement} and \eqref{matrixelementVM} exhibit tensor structures. By compare \eqref{matrixelement} with \eqref{matrixelementVM}, we identify the form factors $G_{1}(q^{2})$ and  $G_{2}(q^{2})$ with elastic form factor $F_{v^n v^n}(q^{2})$
\begin{align}
  \nonumber  G_{1}^{(n)}(q^{2}) &= G_{2}^{(n)}(q^{2}) = F_{v^{n}v^{n}}(q^{2})\,,\\
    G_{3}^{(n)}(q^{2}) &= 0\,.\label{vecelastic}
\end{align}
Note that only two of the three form factors contribute and that both depend only on the elastic form factor. Based on the $G_{C}$, magnetic $G_{M}$, and quadrupole $G_{Q}$ form factors are given by
\begin{align}
     G_{C}^{(n)} &= G_{1}(q^{2}) + \dfrac{q^{2}}{6 m_{v^{m}}^{2}}\bigg\{\left(1 + \dfrac{q^{2}}{4 m_{v^{m}}^{2}}\right)G_{3}(q^{2}) - G_{2}(q^{2})\bigg\},\\
    G_{M}^{(n)} &= G_{1}(q^{2}) + G_{2}(q^{2}),\\
    G_{Q}^{(n)} &= \left(1 + \dfrac{q^{2}}{4 m_{v^{m}}^{2}}\right)G_{3}(q^{2}) - G_{2}(q^{2}).\label{vecformfactors}
\end{align}
 Using the results \eqref{vecelastic}, this last equation allows us to write the electric, magnetic, and quadrupole form factors only in terms of the elastic form factors as
\begin{align}
      G_{C}^{(n)} &= F_{v^{n}v^{n}}(q^{2})\left(1 - \dfrac{q^{2}}{6 m_{v^{m}}^{2}}\right)\label{FE},\\
    G_{M}^{(n)} &= 2\, F_{v^{n}v^{n}}(q^{2})\label{FM},\\
    G_{Q}^{(n)} &= - F_{v^{n}v^{n}}(q^{2})\label{FQ}.
\end{align}
 Figure \ref{fig:electromagneticformfactorsvec}  shows our results for the electromagnetic form factors of vector mesons in the Einstein-dilaton I and II models, in comparison with Lattice QCD \cite{Shultz:2015pfa} (LQCD2015) and the soft-wall model. 

The upper left panel of Fig.~\ref{fig:electromagneticformfactorsvec} displays the electric form factor, while the upper right panel presents the magnetic form factor. The quadrupole form factor is shown in the lower panel. As observed in Fig. \ref{fig:electromagneticformfactorsvec}, the soft-wall model provides the closest overall agreement with the lattice data for the charge form factor. In comparison, Einstein-dilaton model I slightly overestimates, whereas  Einstein-dilaton model II slightly underestimates the electric form factor. For the magnetic form factor, all three models underestimate the lattice QCD data. However, Einstein-dilaton model I provides the closest overall agreement with the lattice results, followed by the soft-wall model, while Einstein-dilaton model II exhibits the largest deviation. Finally, for the quadrupole form factor, all the models predict values systematically below the lattice QCD results over the entire range. Among them, Einstein-dilaton model II yields the closest agreement with the lattice data, followed by the soft-wall model, while Einstein-dilaton model I shows the largest discrepancy. Among the three electromagnetic form factors, the charge form factor exhibits the best agreement with the lattice QCD results. This behavior may be related to the structure of the five-dimensional Yang-Mills action employed to describe the vector sector, which is minimally coupled to the background metric and the dilaton field, as given in Eq.~\eqref{vecaction1}, as well as to the isospin symmetry underlying the electromagnetic structure of the $\rho$ meson.

\medskip

{\bf The ratio between the electromagnetic form factors and the zero of the charge form factor}

\medskip
From the equations \eqref{FE}, \eqref{FM} and \eqref{FQ}, we can obtain the following ratios between the form factors
\begin{align}
    \dfrac{ G_{C}^{(n)}}{G_{M}^{(n)}} = \dfrac{1}{2} - \dfrac{q^{2}}{12 m_{v^{m}}^{2}}\,,\quad  \dfrac{ G_{C}^{(n)}}{G_{Q}^{(n)}} =  -\left(1 - \dfrac{q^{2}}{6 m_{v^{m}}^{2}}\right)\,,\quad \dfrac{ G_{M}^{(n)}}{G_{Q}^{(n)}} = - 2\,. 
\end{align}
According to \cite{Brodsky:1992px}, these ratios are universal and remain valid in both the space-like and time-like regions. Realize that the ratios between the electric and magnetic form factor and between the electric and quadrupole form factor depend on $q^{2}$ and $m_{v^{m}}^{2}$, while the ratio between magnetic and quadrupole form factor is constant. When the ratio $G_{C}/G_{M}$ vanishes, we find $q^{2} = 6 m_{v^{m}}^{2}$ so-called zero of the electric form factor. By substituting this result in \eqref{FE}, one can confirm that indeed electric form factor goes to zero. This is an important and general result within non-perturbative QCD approaches. Using the ground state mass of the $\rho$ vector meson, i.e, $m_{\rho^{0}} = m_{v^{0}} = 0.776\,\text{GeV}$, we find $q^{2} = 3.61\,\text{GeV}^{2}$ for square momentum transferred. This value is consistent with results obtained from the light-front constituent quark model, see, for instance \cite{deMelo:1997hh, Cardarelli:1994yq, deMelo:2018hfw}, where was obtained $q^{2}\approx 4\, \text{GeV}^{2}$. On the other hand, Ref.~\cite{DeMelo:2018bim} reports the relation $q^{2} = 5\,m_{v^{0}}^{2}$, when evaluated using the experimental value of the ground-state vector meson mass, yields $q^{2} \approx 3\,\text{GeV}^{2}$, which is slightly smaller than the value obtained in the present work. Similar value is found in \cite{deMelo:2023zba} using quantum field theory on the light-front. Within the framework of the Dyson-Schwinger equations with contact interaction, the zero is predicted to occur at approximately $q^{2} \approx 5\,\text{GeV}^{2}$, as discussed in Ref.\cite{Roberts:2011wy}. In the present work, we consider the form factors in vacuum. However, the electromagnetic form factors can be sensitive to medium effects. For instance, \cite{deMelo:2018hfw} investigates the electromagnetic form factors of $\rho$ vector mesons in symmetric nuclear matter and reports that the zero of the electric form factor decreases as the nuclear matter density increases. In addition, the zero also exhibits dependence on sub-leading contributions from perturbative QCD amplitudes, as discussed in \cite{deMelo:2016lwr}, where the time-like electromagnetic form factor of the $\rho$ meson was investigated.

\medskip

The form factors can be expanded for high and low $q^{2}$. For large $q^{2}$, the transition and elastic electromagnetic form factors, in the particular case, admit the following expansion 
\begin{align}
 F_{v^n v^k}(q^{2}) =\dfrac{1}{q^{2}}\sum_{m = 0}^{\infty}F_{v^{m}}g_{v^{m}v^{n}v^{k}}\left[1 - \dfrac{m_{v^{m}}^{2}}{q^{2}} + ...\right]\,.\label{largeq2}
\end{align}

{\bf Small $q^{2}$ expansion}

\medskip

On the other hand, in the small $q^{2}$ regime, the electric (or charge), magnetic, and quadrupole form factors admit the following expansion: 
\begin{align}
    G_{C}^{(n)}(q^{2}) &\approx 1 - 2.532 q^{2} + 4.739 q^{4} + \mathcal{O}(q^{6})\,,\\
    G_{M}^{(n)}(q^{2}) &\approx 2 - 4.511 q^{2} + 8.230 q^{4} + \mathcal{O}(q^{6})\,,\\
    G_{Q}^{(n)}(q^{2}) &\approx - 1 + 2.556 q^{2} - 4.115 q^{4} + \mathcal{O}(q^{6})\,,
\end{align}
for model I, while for model II are:
\begin{align}
     G_{C}^{(n)}(q^{2}) &\approx 1 - 2.990 q^{2} + 6.295 q^{4} + \mathcal{O}(q^{6})\,,\\
    G_{M}^{(n)}(q^{2}) &\approx 2 - 5.431 q^{2} + 11.090 q^{4} + \mathcal{O}(q^{6})\,,\\
    G_{Q}^{(n)}(q^{2}) &\approx - 1 + 2.715 q^{2} - 5.545 q^{4} + \mathcal{O}(q^{6})\,.
\end{align}
The electric form factor, in the low $q^{2}$ regime, can be expanded as
\begin{align}
    G_{C}^{(n)} = 1 - \dfrac{1}{6}\langle r_{\rho}^{2}\rangle_{C}\, q^{2} + \dots
\end{align}
where $\langle r_{C}^{2}\rangle$ is the electric (or charge) mean squared radius of the $\rho$ meson, in which is given by
\begin{align}
     \langle r^{2}_{C} \rangle = - 6\, \dfrac{d G_{C}}{dq^{2}}\bigg|_{q^{2} = 0}\,.\label{electricradii}
\end{align}
The same equation \eqref{electricradii} allows us to obtain the corresponding values for the magnetic and quadrupole radii of the vector mesons. In table~\ref{table:VectorMesonElectricRadiusdimensionless}, we summarize our results for the electric, magnetic, and quadrupole dimensionless mean squared radii in the Einstein-dilaton I and II models, compared with lattice QCD data results \cite{Owen:2015gva, Shultz:2015pfa}, as well as with the hard-wall and soft-wall holographic models and the Dyson-Schwinger model (DSM). For completeness, we also express these quantities in the conventional units of fm$^{2}$. The corresponding values for the electric mean squared radius in the hard-wall, soft-wall, and Einstein–dilaton I and II models, expressed in these units, are, respectively,  $0.525$\,\cite{Grigoryan:2007vg}, $0.646$\,\cite{Grigoryan:2007my}, $0.591$ and $0.705$. For the magnetic and quadrupole radii $0.461$, $0.581$, $0.527$ and $0.634$, respectively.  

  The electric mean squared radius of the $\rho$ meson predicted by the Einstein-dilaton II and soft-wall models are closer to the LQCD result \cite{Owen:2015gva}, whereas the model I, hard-wall and DSM models predict larger deviations in comparison to the Lattice QCD data \cite{Owen:2015gva}. The magnetic mean squared radius predicted by the hard-wall model is closer to the result obtained from the Dyson-Schwinger model than those predicted by the Einstein-dilaton I and II, and soft-wall models. The quadrupole mean squared radius predicted by the holographic models differs from the corresponding result obtained in the Dyson-Schwinger model.

\begin{table}[ht!]
\centering
\centering
\footnotesize
\begin{tabular}{l |c|c|c|c|c|c|c}
\hline 
\hline
 & Hard-wall & Soft-wall & Model I & Model II & DSM & LQCD \cite{Owen:2015gva} & LQCD \cite{Shultz:2015pfa}\\
\hline 
$m_{v^0}^{2}\langle r_{C}^{2} \rangle$ & $8.1$\,\cite{Grigoryan:2007vg} & $10$\,\cite{Grigoryan:2007my}  &  $9.1$ & $10.9$ & $7.6$\,\cite{Xu:2024vkn} &  $10.8(1.2)$ &$8.23(15)$\\
\hline
$m_{v^0}^{2}\langle r_{M}^{2} \rangle$ & $7.1$ &  $9$ & $8.1$  & $9.8$ & $6.9$\,\cite{Hernandez-Pinto:2024kwg}  &   &\\
\hline
$m_{v^0}^{2}\langle r_{Q}^{2} \rangle$ & $7.1$  & $9$ &   $8.1$ &  $9.8$ & $5.6$\,\cite{Roberts:2011wy} & &\\
\hline\hline
 \end{tabular}
\caption{Electric, magnetic, and quadrupole dimensionless mean squared radii of vector mesons $\rho$ for Einstein-dilaton models I and II compared against hard-wall, soft-wall, Dyson-Schwinger, and Lattice QCD result.}
\label{table:VectorMesonElectricRadiusdimensionless}
\end{table}

The electric, magnetic, and quadrupole form factors, in the limit $q^{2}\to 0$, allow us to extract the charge, as well the magnetic and quadrupole moments, as
\begin{align}
     G_{C}^{(n)}(0) &= 1\,,\\
     G_{M}^{(n)}(0) =\mu_{n} &= 2\,,\label{magneticmoment}\\
     Q_{n} = \dfrac{1}{m_{v^{m}}^{2}}G_{Q}^{(n)}(0) &= - \dfrac{1}{m_{v^{m}}^{2}}\,,\label{quadrupole}
\end{align}
where $\mu_{n}$ and $Q_{n}$ denote the magnetic and quadrupole moments of the $\rho$ vector meson. In this limit, the electric form factor yields the electric charge of the vector meson. The magnetic moment predicted by the Einstein-dilaton I and II, hard-wall, and soft-wall models is constant. Note that the quadrupole moment in Eq.~\eqref{quadrupole} depends on the square of the vector meson mass. As discussed in \cite{Brodsky:1992px}, massive spin-1 particles without internal substructure are expected to have an electric charge equal to $1$, a magnetic moment equal to $\mu_{n} = 2$, and a quadrupole moment equal to $m_{v^{0}}^{2}Q_{0} = -1$. Therefore, deviations from these values arise once the internal substructure is taken into account, such as the constituents quarks of vector mesons. Deviations in the magnetic and quadrupole moments may also arise when $\rho$ vector mesons are in a nuclear medium; see, for example, \cite{Hutauruk:2025bjd}.

 In tables \ref{table:VectorMesonmagneticmoment} and \ref{table:VectorMesonQuadrupolemoment} we present our results for the magnetic and quadrupole moment of vector mesons in the Einstein-dilaton I and II against Lattice QCD results and Dyson-Schwinger model, hard-wall, and soft-wall. Note that both the Einstein-dilaton I and II models, as well as the hard-wall and soft-wall models, predict the same value for the magnetic moment of the rho meson. This is because our models do not take into account the quarks, and the rho mesons are in a vacuum. The Dyson-Schwinger model, for instance, considers quarks. Our results for the magnetic moment are closer compared to the DSM and LQCD\cite{Shultz:2015pfa}. On the other hand, our result for the quadrupole moment is higher compared to other approaches.

\begin{table}[ht!]
\centering
\centering
\footnotesize
\begin{tabular}{c |c|c|c|c|c|c|c}
\hline 
\hline
 Magnetic & Hard wall & Soft wall & Model I & Model II & DSM & LQCD & LQCD \\
\hline 
$\mu_{n}$ & $2$\,\cite{Grigoryan:2007vg} & $2$\,\cite{Grigoryan:2007my} & $2$  & $2$ & $2.11$\,\cite{Hernandez-Pinto:2024kwg} & $2.61(97)$\,\cite{Owen:2015gva} & $2.17(10)$\cite{Shultz:2015pfa} \\
\hline\hline
 \end{tabular}
\caption{Magnetic moment of vector mesons for Einstein-dilaton models I and II compared against hard-wall, soft-wall, Dyson-Schwinger, and Lattice QCD result. The results are divided by magneton nuclear ($\mu_{N}$).}
\label{table:VectorMesonmagneticmoment}
\end{table}

\begin{table}[ht!]
\centering
\centering
\footnotesize
\begin{tabular}{c |c|c|c|c|c|c|c}
\hline 
\hline
 Quadrupole & Hard-wall & Soft-wall & Model I & Model II & DSM & LQCD & LQCD \\
\hline 
$m_{v^0}^{2}Q_{0}$ & $- 1$ & $- 1$ & $- 1$  & $- 1$ & $- 0.23$\,\cite{Bhagwat:2006pu} & $-0.46$\,\cite{Owen:2015gva} & $-0.54(10)$\cite{Shultz:2015pfa} \\
\hline\hline
 \end{tabular}
\caption{Quadrupole moment of vector mesons for Einstein-dilaton models I and II compared against hard-wall, soft-wall, Dyson-Schwinger, and Lattice QCD result.}
\label{table:VectorMesonQuadrupolemoment}
\end{table}

The form factors satisfy the important sum rule

\begin{align}
    \sum_{m = 0}^{\infty} \dfrac{F_{v^{m}}}{m_{v^{m}}^{2}}g_{v^{m}v^{n}v^{n}} = 1\,.\label{sumrule}
\end{align}

The sum rule \eqref{sumrule} was also found in holographic QCD models based on the top-down approach, for instance Refs.~\cite{BallonBayona:2009ar, Bayona:2011xj, Ballon-Bayona:2012txi}. 


\section{High $q^2$ behavior of the elastic and transition form factors }\label{asympgeneralizedformfactors}

\begin{figure}[htp!]
    \centering
    \begin{subfigure}{0.45\textwidth}
        \centering
        \includegraphics[width=\textwidth]{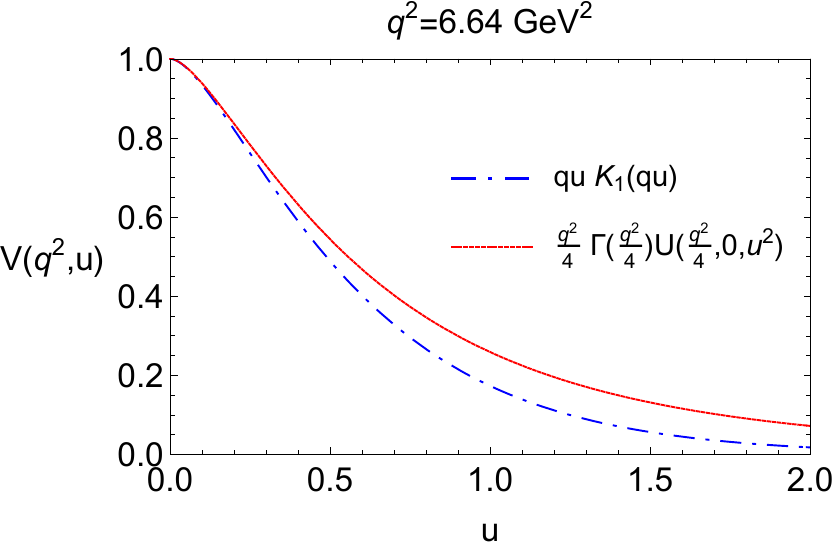}
    \end{subfigure}
    \hspace{0.05\textwidth}
    \begin{subfigure}{0.45\textwidth}
        \centering
        \includegraphics[width=\textwidth]{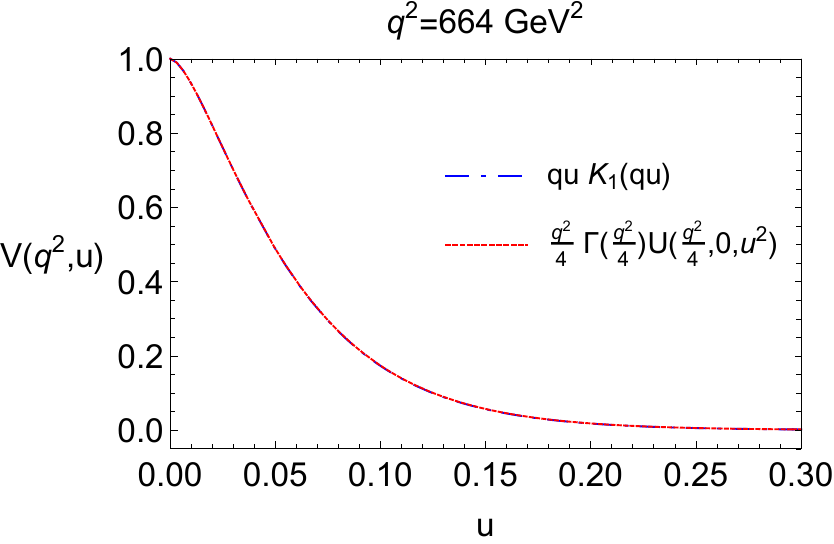}
    \end{subfigure}
    \caption{{\bf Left panel}: Superposition of Eqs~\eqref{Eq:BtBPLargeq} and \eqref{Eq:ASolSW} in the regime of small $q$. {\bf Right panel}: Superposition of Eqs~\eqref{Eq:BtBPLargeq} and \eqref{Eq:ASolSW} in the regime of large $q$.}
    \label{Fig:KU}
\end{figure}

In this section we show details of the behavior of the elastic and transition form factors in the regime of large momentum transfer obtained by solving the integral \eqref{vecformfactorgeneral}. First, we focus on the discussion about the behavior of the bulk to boundary propagator, as highlighted in refs.~\cite{Grigoryan:2007my, Grigoryan:2007vg}, only small values of $u$ are important in the region of large $q$. This means that the leading terms of the differential equation \eqref{bulk to boundary propagator} are
\noindent
\begin{equation}\label{Eq:Bulk-to-Boundary-UV}
    V''(q^2,u)-\frac{1}{u^2}V'(q^2,u)-q^2V(q^2,u)=0.
\end{equation}
\noindent
The solutions of this differential equation are
\noindent
\begin{equation}
    V(q^2,u)=-C_1iu\, I_{1}(q\,u)+C_2 u\,Y_{1}(-iq\,u),
\end{equation}
\noindent
where $C_1$ and $C_2$ are the integration constants, $I_{1}(x)$ is the modified Bessel function of the first kind, and $Y_{1}(x)$ is the Bessel function of the second kind. The integration constant $C_2$ is fixed with the boundary condition $V(q^2,0)=1$, while the other integration constant is fixed to get rid the complex solution. Thus, the solution we are interested in is
\noindent
\begin{equation}\label{Eq:BtBPLargeq}
    V(q^2,u)=q\,u\, K_{1}(q\,u),
\end{equation}
\noindent
where $K_1(q\,u)$ is modified Bessel function of the second kind. In Ref.~\cite{Grigoryan:2007vg}, the authors realized that Eq.~\eqref{Eq:BtBPLargeq} captures the relevant information of the problem in the regime of large $q$. Let us compare this function with the analytic solution obtained in the soft wall model, i.e., Eq.~\eqref{softsolution}, where we used the recurrence relation 
\noindent
\begin{equation}\label{Eq:ASolSW}
    V^{S}(q^2,u)=\frac{q^2}{4}\Gamma\left(\frac{q^2}{4}\right)U\left(\frac{q^2}{4},0,u^2\right).
\end{equation}
\noindent
A superposition of Eqs.~\eqref{Eq:BtBPLargeq} and \eqref{Eq:ASolSW} are displayed in Fig.~\ref{Fig:KU}. Left panel shows the results in the regime of small $q$, where dashed-dot blue line represents Eq.~\eqref{Eq:BtBPLargeq}, while dashed red line represents Eq.~\eqref{Eq:ASolSW}. In turn, the right panel shows the results in the regime of large $q$. As can be seen, both functions seems to be the same in the region of large $q$. To investigate if these functions are really equivalent we calculated the difference, i.e., $\Delta V=q\,u\, K_{1}(q\,u)-V^{S}(q^2,u)$; for $q^2=664$  GeV$^2$. Our numerical results show that $\Delta V$ lies within the interval $[-10^{-3},10^{-3}]$ over the domain of integration. These results support the statement that Eqs.~ \eqref{Eq:BtBPLargeq}  and \eqref{Eq:ASolSW} are equivalent in the regime of large $q$.

\begin{figure}[htp!]
    \centering
    \begin{subfigure}{0.45\textwidth}
        \centering
        \includegraphics[width=\textwidth]{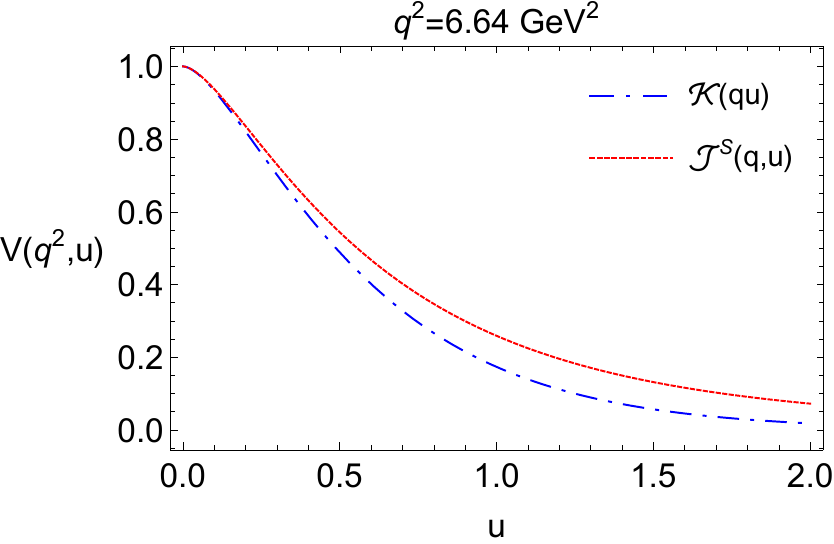}
    \end{subfigure}
    \hspace{0.05\textwidth}
    \begin{subfigure}{0.45\textwidth}
        \centering
        \includegraphics[width=\textwidth]{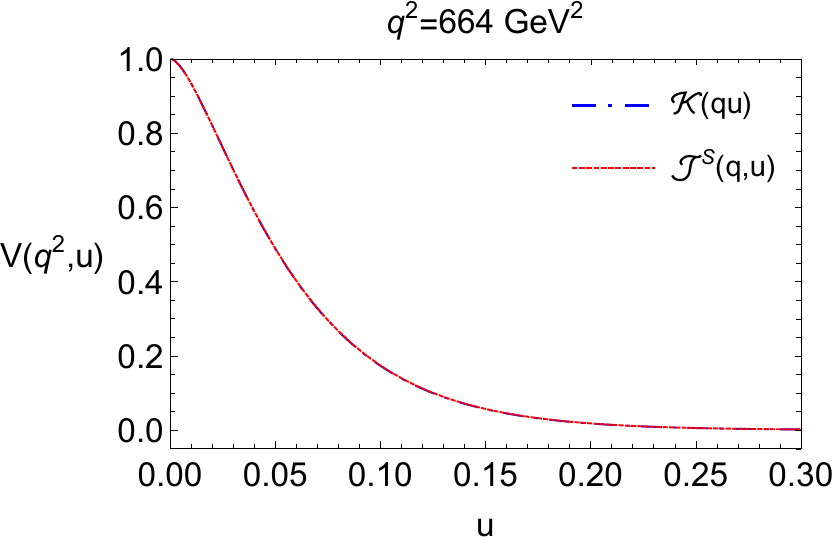}
    \end{subfigure}
    \caption{{\bf Left panel}: Superposition of Eqs~\eqref{Eq:K} and \eqref{Eq:J} in the regime of small-$q$. {\bf Right panel}: Superposition of Eqs~\eqref{Eq:K} and \eqref{Eq:J} in the regime of large $q$.}
    \label{Fig:KJ}
\end{figure}

An alternative procedure to investigate the behavior of large $q$ is considering the integral representation of the corresponding equations \eqref{Eq:BtBPLargeq} and \eqref{Eq:ASolSW}. In Ref.~\cite{Grigoryan:2007vg}, the authors wrote the equation $\mathcal{K}(q\,u)\equiv qu K_{1}(q\,u)$, where $\mathcal{K}(q\,u)$ has an integral representation given by
\noindent
\begin{equation}\label{Eq:K}
    \mathcal{K}(q\,u)=u^2\int_{0}^{1}\frac{dx}{(1-x)^2}\exp{\left[-\frac{(1-x)q^2}{4x}-\frac{x\,u^2}{1-x}\right]}.
\end{equation}
\noindent
In turn, the integral representation of Eq.~\eqref{Eq:ASolSW} is obtained after an integration by parts of Eq.~\eqref{btbp}, 
\noindent 
\begin{equation}\label{Eq:J}
    \mathcal{J}^{S}(q,u)=u^2\int_{0}^{1}\frac{dx}{(1-x)^2}\exp{\left[-\frac{q^2}{4}\ln{\left(\frac{1}{x}\right)}-\frac{x\,u^2}{1-x}\right]}.
\end{equation}
\noindent
Thus, both equations \eqref{Eq:K} and \eqref{Eq:J} must be equivalent in the regime of large $q$. To better visualize this equivalence we plot the superposition of the integral representations. We display our results for small $q$ in the left panel of Fig.~\ref{Fig:KJ}, while the right panel shows the results for large $q$. As can be seen, both functions are equivalent in the regime of large $q$. As before, we also compute the difference $\Delta V=\mathcal{K}-\mathcal{J}^{S}$, which lies within the interval $[-10^{-3},10^{-3}]$ for $q^2=664$ GeV$^2$. It is worth pointing out that the relevant contribution of the bulk to boundary propagator is restricted to the region close to the boundary in the regime of large $q$, see Appendix \ref{vecbtbp} for details.

\noindent

\begin{figure}[htp!]
    \centering
    \begin{subfigure}{0.45\textwidth}
        \centering
        \includegraphics[width=\textwidth]{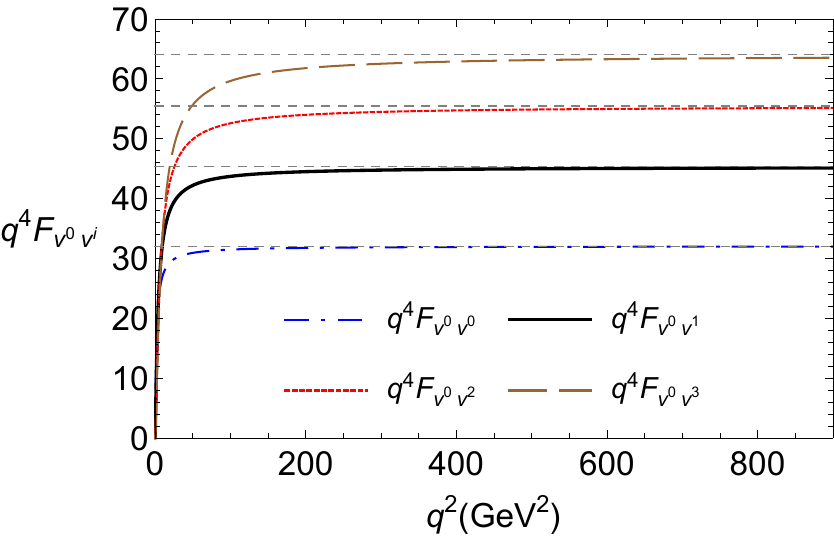}
    \end{subfigure}
    \hspace{0.05\textwidth}
    \begin{subfigure}{0.45\textwidth}
        \centering
        \includegraphics[width=\textwidth]{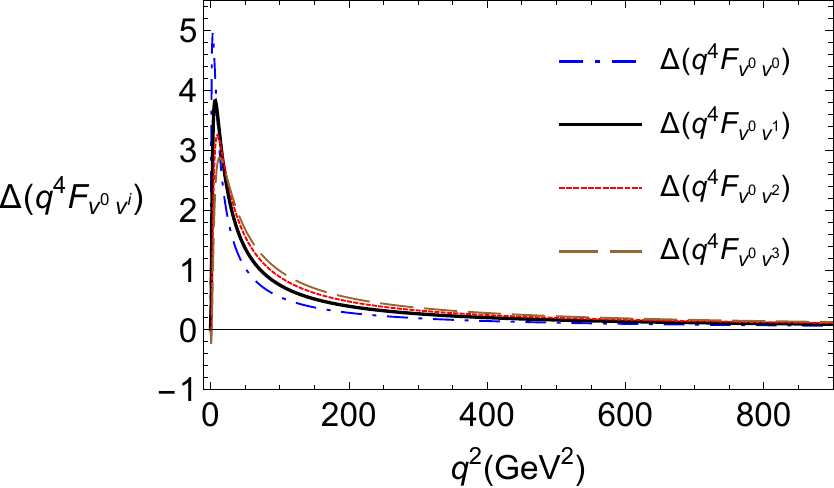}
    \end{subfigure}
    \caption{{\bf Left panel}: Elastic and transition form factors of vector mesons for the soft wall model. {\bf Right panel}: The difference of the propagators given by Eq.~\eqref{Eq:DeltaF0i}.}
    \label{Plot:FmnSW}
\end{figure}

On the other hand, with the results obtained above we can investigate the asymptotic behavior of the elastic and transition form factors in the regime of large $q$ considering the fact that in this regime the bulk to boundary propagator can be approximated by Eq.~\eqref{Eq:BtBPLargeq}. In turn, as pointed out in Refs.~\cite{Grigoryan:2007my, Grigoryan:2007vg} (see also references there in) the relevant contribution of the wave functions is restricted to the asymptotic behavior close to the boundary; for the soft-wall model it is given by
\noindent
\begin{equation}\label{Eq:SWAsympWF}
    v^n(u)=\sqrt{2(n+1)}\,u^2,\qquad u\to 0 \, .
\end{equation}
Comparing against the general asymptotic behavior given by Eq.~\eqref{Eq:AsympBWF}, we conclude that for the soft-wall model the normalization constant is given by $N_{v^n}=\sqrt{2(n+1)}$. Plugging \eqref{Eq:SWAsympWF} and \eqref{Eq:BtBPLargeq} in Eq.~\eqref{vecformfactorgeneral}, then, expanding the result in the regime of large $q$, we obtain the asymptotic behavior
\noindent
\begin{equation}\label{Eq:AsymptBFmn}
    F_{v^{m}v^{n}}(q)=\frac{32\sqrt{1+m}\sqrt{1+n}}{q^4}+\mathcal{O}\left(q^{-6},q^{-8},\cdots\right),\qquad q\to\infty 
\end{equation}
\noindent
where $\mathcal{O}\left(q^{-6},q^{-8},\cdots\right)$ are subleading contributions. Now, we can verify if the asymptotic behavior of the generalized form factors is indeed described by Eq.~\eqref{Eq:AsymptBFmn}. In the left panel of Fig.~\ref{Plot:FmnSW}, we display our numerical results of the elastic and transition form factors of the soft-wall model solving directly the differential equation \eqref{bulk to boundary propagator}, see Appendix \ref{vecbtbp}. As can be seen, the elastic and transition form factors approach asymptotically to the horizontal dashed lines representing the asymptotic behavior given by Eq.~\eqref{Eq:AsymptBFmn}. In turn, to see the region where we can use $\mathcal{K}(qu)$ instead of $\mathcal{J}^{S}(q,u)$ we can plot the difference of the elastic and transition form factors involving these functions,
\noindent
\begin{equation}\label{Eq:DeltaF0i}
    \Delta \left(q^4F_{v^{0}v^{i}}\right)=q^4\int_{0}^{\infty}du\frac{e^{-u^2}}{u}\left(\mathcal{J}^{S}(q,u)-\mathcal{K}(qu)\right)v^{0}(u)v^{i}(u).
\end{equation}
\noindent
We display the results in the right panel of Fig.~\ref{Plot:FmnSW}. 
As can be seen, the functions $\mathcal{J}^{S}(q,u)$, and $\mathcal{K}(qu)$ provide different values in the region of small momentum transfer, while they provide the same results in the region of large momentum transfer. This result is in agreement with the assumption that $\mathcal{K}(qu)$  can be used instead of $\mathcal{J}^{S}(q,u)$ in the region of large momentum transfer.

\noindent

\begin{figure}[htp!]
    \centering
    \begin{subfigure}{0.45\textwidth}
        \centering
        \includegraphics[width=\textwidth]{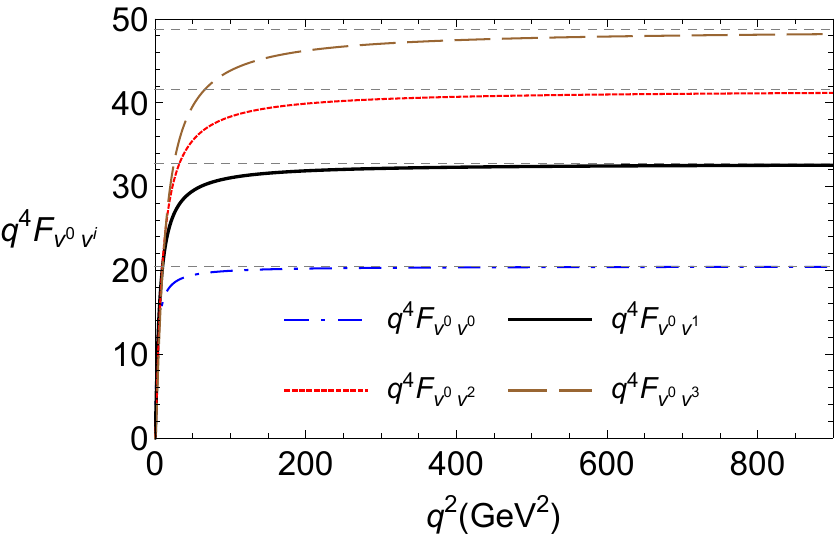}
    \end{subfigure}
    \hspace{0.05\textwidth}
    \begin{subfigure}{0.45\textwidth}
        \centering
        \includegraphics[width=\textwidth]{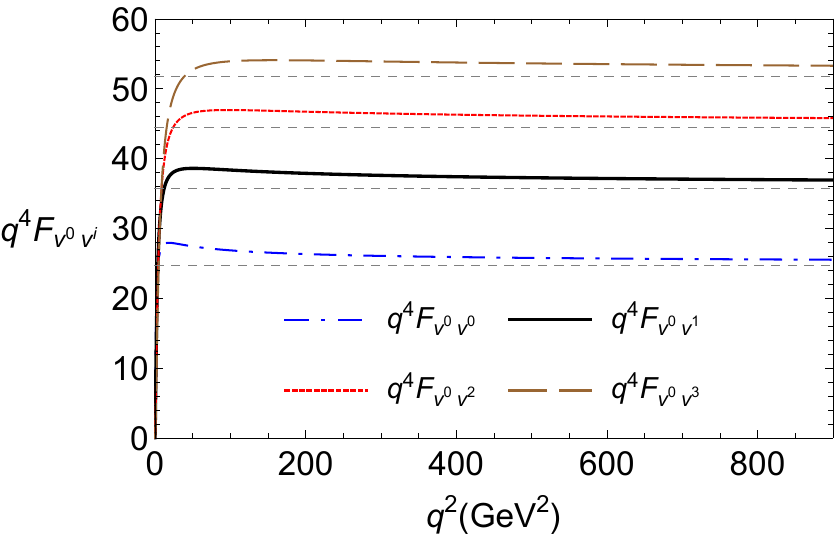}
    \end{subfigure}
    \caption{{\bf Left panel}: Elastic and transition form factors of vector mesons for model I. {\bf Right panel}: Elastic and transition form factors of vector mesons for model II}
    \label{Plot:FmnMIMII}
\end{figure}

On the other hand, we can extend the previous analysis for the Einstein-dilaton models. From the analysis developed in Appendix \ref{vecbtbp}, we conclude that the bulk to boundary propagator of the Einstein-dilaton models I and II are equivalent to the bulk to boundary propagator of the the soft-wall model, thus, it can be approximated by Eq.~\eqref{Eq:BtBPLargeq} in the regime of large $q$. In turn, the asymptotic behavior of the wave functions of the Einstein-dilaton models is given by Eq.~\eqref{Eq:AsympBWF}. Plugging these information in Eq.~\eqref{vecformfactorgeneral}, then, expanding the result in the limit of large $q$, we obtain 
\noindent
\begin{equation}\label{Eq:EDAsymptBFmn}
    F_{v^{m}v^{n}}(q)=\frac{16N_{v^{m}}N_{v^{n}}}{q^4}+\mathcal{O}\left(q^{-6},q^{-8},\cdots\right),\qquad q\to\infty 
\end{equation}
\noindent
where $N_{v^{m}}$ and $N_{v^{n}}$ are the normalization constants. Notice that Eq.~\eqref{Eq:AsymptBFmn} is a particular case of the general result \eqref{Eq:EDAsymptBFmn}. It is worth pointing out that we can rewrite the asymptotic behavior of the elastic and transition form factors in terms of the decay constants using the relation \eqref{2pointvecdecay} 
\noindent
\begin{equation}\label{Eq:FmnLq}
    F_{v^{m}v^{n}}(q)=\frac{4g_5^2F_{v^{m}}F_{v^{n}}}{q^4}+\mathcal{O}\left(q^{-6},q^{-8},\cdots\right),\qquad q\to\infty 
\end{equation}
\noindent
where $g_5^2=12\pi^2/N_c$ (in our case we use $N_c=3$). This result represent our prevision of the asymptotic behavior of the elastic and transition form factors in the regime of large $q$. For completeness, we also solve numerically the differential equation for models I and II. In the left panel of Fig.~\ref{Plot:FmnMIMII}, we display our numerical results of the elastic and transition form factors for model I. As can be seen, the elastic and transition form factors approach to the asymptotic values in the regime of large $q$ by bellow. The corresponding results for model II are displayed in the right panel of Fig.~\ref{Plot:FmnMIMII}, where the results approach to the asymptotic values by above.

\noindent

To finish this section we can obtain additional information by comparing the elastic and transition form factors obtained through Eqs.~\eqref{Eq:FmnLq} and \eqref{largeq2} in the limit of large $q^2$. For vector mesons, it is expected that the form factor for large $q^{2}$ behaves as $ F_{v^{n}v^{k}}(q^{2})\sim 1/q^{4}$ \cite{Brodsky:2007hb, Hong:2004sa}. Thus, We can write Eq.~\eqref{largeq2} in the form 
\noindent
\begin{align}
 F_{v^n v^k}(q^{2}) =\dfrac{1}{q^{2}}\sum_{m = 0}^{\infty}F_{v^{m}}g_{v^{m}v^{n}v^{k}} - \frac{1}{q^4}\sum_{m = 0}^{\infty}F_{v^{m}}g_{v^{m}v^{n}v^{k}}m_{v^{m}}^{2} + \dots .\label{largeq2n2}
\end{align}
\noindent
By comparing this result against \eqref{Eq:FmnLq}, we conclude that the coefficient of $1/q^2$ must satisfy the relation
\noindent
\begin{align}
    \sum_{m = 0}^{\infty} F_{v^{m}}g_{v^{m}v^{n}v^{k}} = 0\,.\label{superconv}
\end{align}
\noindent
This is the superconvergence relation. We can also calculate this superconvergence relation using the vector meson dominance representation, i.e., Eq.~\eqref{largeq2}, considering the first ten values of the decay constants and couplings displayed in tables \ref{Table:VectorMesonDecayConstants} and \ref{table:VectorMesonCouplings}, for the Einstein dilaton models I and II, we have
\begin{align}
    \sum_{m = 0}^{9} F_{v^{m}}g_{v^{m}v^{0}v^{0}} &\approx - 0.000178245,\qquad (\text{model I})\label{verifysuperconvI}\\
    \sum_{m = 0}^{9} F_{v^{m}}g_{v^{m}v^{0}v^{0}} &\approx - 0.0814468.\quad \qquad (\text{model II})\label{verifysuperconvII}
\end{align}
It is expected that the value of this sum shall decrease, converging to zero, as we add more eigenstates \cite{BallonBayona:2009ar}. Therefore, the results \eqref{verifysuperconvI} and \eqref{verifysuperconvII} show that our models satisfy the superconvergence relation \eqref{superconv}. Meanwhile, by comparing the coefficients of $1/q^4$, we obtain the sum rule
\noindent
\begin{equation}
    \sum_{m = 0}^{\infty}F_{v^{m}}g_{v^{m}v^{n}v^{k}}m_{v^{m}}^{2}=-4g_5^2F_{v^{n}}F_{v^{k}}.\label{Eq:SumRuleN2}
\end{equation}
\noindent
This result relates the decay constants, couplings and masses. Similar expressions were obtained in other holographic models, see for instance \cite{Sakai:2005yt, Hirn:2005nr}. We verified that Eq.~\eqref{Eq:SumRuleN2} is satisfied for the results obtained in the soft-wall model in the elastic and transition cases. In turn, for the model I, the sum rule is satisfied up to the ``first decimal place'' with ten eigenvalues in the elastic case; while for the transition case one need more eigenvalues. For the model II, the convergence is poor, one need to consider more that ten eigenvalues even in the elastic case. This result can be understand by observing the behavior of the transition form factors displayed in Fig.~\ref{Plot:FmnMIMII}.

\section{Conclusions}\label{section4}


By employing Einstein–dilaton holographic QCD models, we analyzed the electromagnetic structure of vector mesons through two independent formulations based on three-point correlation functions and a Kaluza–Klein approach, which yield consistent results. This framework enables a consistent treatment of both diagonal and non-diagonal matrix elements of the electromagnetic current.

The resulting charge, magnetic, and quadrupole form factors show different levels of agreement with the available lattice QCD data for the $\rho$ meson. In particular, the charge form factor shows the best agreement with the lattice data, whereas the magnetic and quadrupole form factors are systematically underestimated, with model I providing the closest description of the magnetic form factor and model II yielding the best agreement for the quadrupole form factor. The generalized form factors are described within a vector meson dominance framework, with strong couplings governed by overlap integrals of normalizable modes. We find an approximate selection rule for transitions involving vector meson states $v^{n}$ and $v^{k}$, that becomes exact in the soft wall model, given by $v^{n}$ and $v^{k}$ satisfying $|m - k| \le 1$. This leads to a hierarchy in which transitions between nearby states are enhanced, while those involving more distant excitations are suppressed.

At large momentum transfer, the form factors display the expected QCD scaling behavior, with the asymptotic normalization fixed within the holographic framework. We also identify a nontrivial sum rule connecting masses, decay constants, and couplings, together with the fulfillment of superconvergence constraints. Furthermore, we show that for any AdS/QCD model with the vector sector been described by a five-dimensional Yang-Mills action, the electric form factor exhibits a zero at $q^{2} = 6\, m_{\rho^{0}}^{2} = 3.61\,\text{GeV}^{2}$, in agreement with light-front and Dyson–Schwinger results.

These results support the use of Einstein–dilaton models as a consistent setting for describing both elastic and transition observables of vector mesons, including excited states. More broadly, they indicate that holographic constructions of this type can capture key aspects of nonperturbative QCD dynamics, particularly in regimes where lattice calculations remain unavailable.

Natural extensions of this work include the study of other hadronic channels as nucleons based on the model constructed in \cite{Ballon-Bayona:2024yuz}, as well as the investigation of medium effects such as finite temperature and density \cite{dePaula:2020bte}. A further direction is the comparison with future lattice QCD results for transition form factors, which will provide a more stringent test of the present framework. Future directions also include the investigation of helicity amplitudes and timelike electromagnetic form factors of the $\rho$ meson, following the approach of Ref.\cite{Dbeyssi:2011ep}. Another interesting extension is the analysis of the corresponding structure functions within the same framework, along the lines of Ref.\cite{BallonBayona:2010ae}. Another possible extension would be to include chiral and flavor symmetry breaking in the electromagnetic form factors of charmed mesons, generalizing the works done in \cite{Ballon-Bayona:2017bwk, Ahmed:2023zkk}. In addition, it would be interesting to investigate the gravitational form factors of vector mesons and nucleons based on \cite{Abidin:2008ku, Abidin:2008hn, Abidin:2009hr}. Finally, a natural extension of the present work would be to incorporate higher-dimensional operators into the five-dimensional action, following the approach of Ref.~\cite{Grigoryan:2007iy}, and investigate their impact on the electromagnetic structure of the vector mesons.
These projects will be developed in the future.

\acknowledgments

The authors thank J. J. Dudek for discussions during the development of this work and for providing the lattice QCD data on the electromagnetic form factors of $\rho$ mesons of Ref.\cite{Shultz:2015pfa}. The work of the author A.B-B was partially funded by Conselho Nacional de Desenvolvimento Cient\'\i fico e Tecnol\'ogico (CNPq, Brazil), Grant No. 314000/2021-6, and Coordena\c{c}\~ao de Aperfei\c{c}oamento do Pessoal de N\'ivel Superior (CAPES, Brazil), Finance Code 001. A.S.S. Jr  acknowledges support from “Fundação Carlos Chagas Filho de Amparo à Pesquisa do Estado do Rio de Janeiro” – FAPERJ, Processo SEI-260003/013507/2025.  TF thanks the the support from CNPq (Grant No. 306834/2022-7) and FAPESP (Grant No. 2023/13749-1). WdP acknowledges the partial support of the CNPq (Grants No. 3313030/2021-9, 401565/2023-8,408419/2024-5) and FAPESP (Grant 2025/05312-8).
\appendix

\section{The Kaluza-Klein expansion}\label{App:kkexpansion}

Here we present an alternative procedure to obtain the electromagnetic form factors of vector mesons. We follow the procedure presented in  Refs.~\cite{Grigoryan:2007my,Grigoryan:2007vg,Ballon-Bayona:2017bwk} adapted for the problem we are dealing with in this paper. The goal is to expand the action \eqref{vecaction1} up to third-order in the field $V^{a}_{m}$. We start by plugging Eq.~\eqref{Eq:FieldStrength} in \eqref{vecaction1}, then, considering the action containing second-order terms in $V^{a}_{m}$,
\noindent
\begin{equation}\label{Eq:s2Action}
        S^{(2)}=-\int d^5x\sqrt{|g|}e^{-\Phi}\Big\{\frac{1}{4g_5^2}v^{mn\,a}v_{mn}^a\Big\},
\end{equation}
\noindent
where $v_{mn}=\partial_mV_n-\partial_nV_m$. The variation of this action can be written as
\begin{equation*}
    \delta S^{(2)}=
    \int d^5x\,\partial_m\left(P_V^{mn\,a} \delta V_{n}^a\right)-\int d^5x\,\partial_m\left(P_V^{mn\,a}\right)\delta V_{n}^a,
\end{equation*}
where we have defined the conjugate momenta
\begin{equation}
    P_V^{mn\,a}:=\frac{\partial\mathcal{L}^{(2)}}{\partial (\partial_mV^{a}_n)}=-\sqrt{|g|}\frac{e^{-\Phi}}{g_5^2}v^{mn\,a}.
\end{equation}
The second term of $\delta S^{(2)}$ leads to the equation of motion for an arbitrary $\delta V_{n}^a$
\noindent
\begin{equation}
    \partial_m\left(\sqrt{|g|}e^{-\Phi}v^{mn\,a}\right)=0,
\end{equation}
\noindent
while the boundary term can be written as
\noindent
\begin{equation*}
        \delta S^{(2),\,\text{bdy}}=
        \int d^4x\int_{\epsilon}^{\infty}du\,\left[\partial_u\left(P_V^{un\,a} \delta V_{n}^a\right)+\partial_\mu\left(P_V^{\mu n\,a} \delta V_{n}^a\right)\right].
\end{equation*}
\noindent
Imposing periodic boundary conditions in the $x^{\mu}$ coordinates the second term of the last equation vanishes, then, the boundary term becomes
\noindent
\begin{equation*}
        \delta S^{(2),\,\text{bdy}}
        =-\int d^4x\left[P_V^{u\mu\,a} \delta V_{\mu}^a\right]_{\epsilon}=-\int {d}^4{x}\langle{J}^{\mu\,a}_V\rangle\left(\delta V_{\mu}^a\right)_\epsilon
\end{equation*}
\noindent
where the four-dimensional current is given by
\noindent
\begin{equation}\label{Eq:VectorCurrent}
    \langle{J}^{\hat\mu\,a}_V\rangle=-\frac{e^{A_s-\Phi}}{g_5^2}\partial_uV^{\hat\mu\,a}\bigg{|}_{u=\epsilon},
\end{equation}
\noindent
here $\hat\mu$ represents the Minkowski index, and we use the fact that $V_u=0$. 

In turn, we can decompose the integral and field in the form, $dx^5=dudx^4$ and $V_m=\{V_u,V_\mu\}$. Plugging these decompositions in \eqref{Eq:s2Action}, the result can be written in terms of the boundary indices $V_{\hat\mu}$. Thus, the second-order action becomes
\noindent
\begin{equation}\label{Eq:ActionVectorSec}
    S^{(2)}=\int d^4x\int du\, e^{A_s-\Phi}\Big\{-\frac{1}{4g_5^2}\left[\left(v_{\hat\mu \hat\nu}^{a}\right)^2+2\left(v_{u\hat\nu}^{a}\right)^2\right]\Big\}.
\end{equation}
\noindent
Next, the idea is to write this action as a four-dimensional effective action of a Lagrangian density, for doing that we write
\noindent
\begin{equation}\label{Eq:GaugeSymmVector}
    \begin{split}
    V_{\hat\mu}^a=\,&V_{\hat\mu}^{\perp, a},
    \end{split}
\end{equation}
\noindent
where the transverse vector satisfy the condition $\partial_{\hat\mu}V_{\hat\mu}^{\perp,a}=0$; the five-dimensional field $V_{\hat\mu}^{\perp, a}$ describes an infinite tower of four-dimensional massive spin $1$ fields. Replacing Eq.~\eqref{Eq:GaugeSymmVector} into the action \eqref{Eq:ActionVectorSec}, we obtain
\begin{equation}\label{Eq:ActionVectorSecN2}
    \begin{split}
    S^{(2)}
    =\,&\int d^4x\int du\, e^{A_s-\Phi}\Big\{-\frac{1}{4g_5^2}\Big[\left(v_{\hat\mu \hat\nu}^{\perp,\,a}\right)^2+2\left(\partial_{u}V_{\hat\nu}^{\perp,\,a}\right)^2\Big]\Big\}.
    \end{split}
\end{equation}
\noindent
The next step is to consider the Kaluza-Klein expansion of the field:
\noindent
\begin{equation}\label{Eq:kkexpansionVector}
    \begin{split}
    V_{\hat\mu}^{\perp,a}(x,u)=\,&g_5\sum_{n=0}^{\infty}v^{a,n}(u)\hat{V}_{\hat\mu}^{a,n}(x).
    \end{split}
\end{equation}
Plugging \eqref{Eq:kkexpansionVector} in \eqref{Eq:ActionVectorSecN2}, the action can be written as
\noindent
\begin{equation*}
    \begin{split}
        S^{(2)}
        =\,&\int d^4x\,\mathcal{L}_V,
    \end{split}
\end{equation*}
where 
\begin{equation}\label{Eq:LagrangianVector}
    \mathcal{L}_V=-\frac{1}{4}\Delta_V^{a,\,nm}\hat{v}^{\hat\mu\hat\nu\,a,n}(x)\hat{v}_{\hat\mu\hat\nu}^{a,m}(x)+\frac{1}{2}M_V^{a,nm}\hat{V}^{\hat\mu\,a,n}(x)\hat{V}_{\hat\mu}^{a,m}(x),
\end{equation}
and 
\begin{equation*}
    \begin{split}
        \Delta_V^{a,nm}=\,&\int du\, e^{A_s - \Phi}v^{a,n}(u)v^{a,m}(u),\\
        M_V^{a,nm}=\,&\int du\, e^{A_s - \Phi}\partial_uv^{a,n}(u)\partial_uv^{a,m}(u).
    \end{split}
\end{equation*}
\noindent
In order to obtain the standard kinetic terms in Eqs.~\eqref{Eq:LagrangianVector}, we consider \cite{Ballon-Bayona:2017bwk}
\begin{equation}
    \begin{split}
    \Delta_V^{a,nm}=\,&\delta^{nm}\\
    M_V^{a,nm}=\,&m_{V^{a,n}}^2\delta^{nm}.
    \end{split}
\end{equation}
Plugging these expressions into Eq.~\eqref{Eq:LagrangianVector}, we obtain the effective Lagrangian density
\noindent
\begin{equation}\label{Eq:LagrangianVectorN2}
    \mathcal{L}_V=-\frac{1}{4}\hat{v}^{\hat\mu\nu\,a,n}\hat{v}_{\hat\mu\nu}^{a,n}-\frac{1}{2}m_{V^{a,n}}^2\hat{V}^{\hat\mu\,a,n}\hat{V}_{\hat\mu}^{a,n}.
\end{equation}
\noindent

On the other hand, plugging Eqs.~\eqref{Eq:GaugeSymmVector} and  \eqref{Eq:kkexpansionVector} into the four-dimensional current, \eqref{Eq:VectorCurrent}, it becomes
\noindent
\begin{equation}\label{Eq:VectorCurent}
    \langle J^{\hat\mu\,a}_V\rangle=\sum_{n=1}^{\infty}g_{V^{a,n}}\hat{V}^{\hat\mu\,a,n}(x),
\end{equation}
\noindent
where we have introduced the leptonic decay constant defined by
\noindent
\begin{equation}
    g_{V^{a,n}}=\frac{e^{A_s - \Phi}}{g_5}\partial_uv^{a,n}(u)\bigg{|}_{u=\epsilon}.
\end{equation}
\noindent
Taking the divergence of Eq.~\eqref{Eq:VectorCurent}, and considering that $\partial_{\hat\mu}V^{\hat\mu\,\perp,a}=\partial_{\hat\mu}\hat{V}^{\hat\mu\,a,n}(x)=0$,
\noindent
\begin{equation}\label{Eq:DivergenceVectorCurrent}
    \begin{split}
        \partial_{\hat\mu}\langle J^{\hat\mu\,a}_V\rangle=0.
    \end{split}
\end{equation}
Indicating that the current is conserved. In turn, the third-order action is given by
\noindent
\begin{equation}\label{Eq:ThirdOrderAction}
    \begin{split}
        S^{(3)}=\,& -\int d^5x\sqrt{|g|}e^{-\Phi}\Big\{\frac{1}{2g_5^2}f^{abc}v^{mn\,a}V_m^bV_n^c\Big\}.
    \end{split}
\end{equation}
\noindent
Repeating the procedure outlined above, we obtain
\noindent
\begin{equation}\label{Eq:Svvv2}
    S^{(3)}=-\frac{1}{2g_5^2}f^{abc}\int dx^4\int du\, e^{A_s-\Phi}v^{\hat\mu \hat\nu\,a}_{\perp}V_{\hat\mu}^{\perp,b}V_{\hat\nu}^{\perp, c},
\end{equation}
\noindent
where $v_{\hat\mu \hat\nu}^{\perp\,a}=\partial_{\hat\mu}\hat{V}_{\hat\nu}^{\perp\,a}-\partial_{\hat\nu}\hat{V}_{\hat\mu}^{\perp\,a}$, plugging the decomposition \eqref{Eq:kkexpansionVector} into \eqref{Eq:Svvv2}, the action can be written in the form
\noindent
\begin{equation}\label{Eq:Svvv3Simpl}
        S^{(3)}
        =\int dx^4 \mathcal{L}_{VVV}
\end{equation}
\noindent
where
\noindent
\begin{equation}\label{Eq:Lvvv}
    \mathcal{L}_{VVV}=f^{abc}g_{\hat{V}^{a,\ell}\hat{V}^{b,m}\hat{V}^{c,n}} \hat{V}^{\hat\mu\, a,m}\left(\partial_{\hat\mu}\hat{V}_{\hat\nu}^{b,\ell}\right)\hat{V}^{\hat\nu\, c,n}
\end{equation}
\noindent
and
\noindent
\begin{equation}\label{Eq:SvvvCouplingN2}
    g_{\hat{V}^{a,\ell}\hat{V}^{b,m}\hat{V}^{c,n}}=g_5 \int du\, e^{A_s-\Phi}\, v^{a,\ell}(u)v^{b,m}(u)v^{c,n}(u).
\end{equation}
\noindent
Here it is worth pointing out that Eq.~\eqref{Eq:SvvvCouplingN2} is the same as Eq.~\eqref{couplingvecnk} \footnote{Note that our couplings are related to the couplings defined in Ref. \cite{Ballon-Bayona:2017bwk} by a factor of $2$.}. We have used both equations to compute the couplings and obtained the same results. The result obtained here, i.e., Eq.~\eqref{Eq:SvvvCouplingN2}, shows the consistency of the procedure outlined in this Appendix as an alternative method to calculate the coupling between three vector mesons. One can go one step further and use the Feynman rules to obtain the matrix element of the interaction represented in the Feynman diagram displayed in Fig.~\ref{Plot:FeynmannVM}, where $v^{k}$ is the initial vector meson state with momenta $p$ and polarization $\epsilon$, and $v^{n}$ is the final vector meson state with momenta $p'$ and polarization $\epsilon'$, while the momenta of the incoming photon is $q$. The conservation of momentum implies that $p'=p+q$. Meanwhile, the electromagnetic current is given by
\noindent
\begin{equation}
    \langle J^{\hat \mu\,a}(x)\rangle=g_{V^{a,m}}\hat{V}^{\hat\mu\,a,m},
\end{equation}
\noindent
where $g_{V^{a,\,m}}$ is the decay constant. The current is the term that will interact with the incoming vector meson $v^{k}$ exchanging an off-shell vector state. From the effective Lagrangian \eqref{Eq:LagrangianVectorN2} and the interaction \eqref{Eq:Svvv3Simpl}, we find the matrix element
\begin{equation}\label{Eq:MatrizElement}
    \begin{split}
    \langle v^{a,k}(p),\epsilon|J^{\hat\mu\,c}(0)|v^{b,n}(p'),\epsilon'\rangle=\,&\epsilon^{\hat\nu}{\epsilon'}^{\hat\rho}f^{abc}\left[\eta_{\hat\sigma\hat\nu}\left(q-p\right)_{\hat\rho}+\eta_{\hat\nu\hat\rho}\left(2p+q\right)_{\hat\sigma}-\eta_{\hat\rho\hat\sigma}\left(p+2q\right)_{\hat\nu}\right]\\
    &\times g_{\hat{V}^{a,m}}g_{\hat{V}^{b,k}\hat{V}^{c,n}\hat{V}^{d,n}}\left[\frac{\eta^{\hat\mu\hat\sigma}+\frac{q^{\hat\mu}q^{\hat\sigma}}{M_{\hat{V}^{a,m}}^2}}{q^2+M_{\hat{V}^{a,m}}^2}\right].
    \end{split}
\end{equation}
\noindent
Considering the sum rule 
\noindent
\begin{equation}\label{Eq:SumRule}
    \sum_{m=0}^{\infty}\frac{g_{\hat{V}^{b,k}\hat{V}^{c,m}\hat{V}^{d,n}}}{M_{\hat{V}^{a,m}}^2}=\delta_{kn},
\end{equation}
\noindent
Eq.~\eqref{Eq:MatrizElement} can be written in the form 
\noindent
\begin{equation}\label{Eq:MatrizElementN2}
    \begin{split}
    \langle v^{a,k}(p),\epsilon|J^{\hat\mu\,c}(0)|v^{b,n}(p'),\epsilon'\rangle&=\epsilon^{\hat\nu}{\epsilon'}^{\hat\rho}f^{abc}\left[\eta_{\hat\sigma\hat\nu}\left(q-p\right)_{\hat\rho}+\eta_{\hat\nu\hat\rho}\left(2p+q\right)_{\hat\sigma}-\eta_{\hat\rho\hat\sigma}\left(p+2q\right)_{\hat\nu}\right]\\
    &\times\left[\left[\eta^{\hat\mu\hat\sigma}-\frac{q^{\hat\mu}q^{\hat\sigma}}{q^2}\right]F_{v^{k}v^{n}}(q^2)+\delta_{kn}\frac{q^{\hat\mu}q^{\hat\sigma}}{q^2}\right],
    \end{split}
\end{equation}
\noindent
where we have introduced the generalized vector meson form factor defined by
\noindent
\begin{equation}\label{Eq:GenFormFactor}
    F_{v^{k}v^{n}}(q^2)=\sum_{m=0}^{\infty}\frac{g_{\hat{V}^{a,m}}g_{\hat{V}^{a,k}\hat{V}^{b,m}\hat{V}^{c,n}}}{q^2+M_{\hat{V}^{a,m}}^2}.
\end{equation}
\noindent

Taking into consideration the transversality of the vector mesons polarizations: $\epsilon^\mu p_{\mu}=0$ and ${\epsilon'}^\nu p'_{\nu}=0$, the matrix element becomes
\noindent
\begin{equation}\label{Eq:MatrizElementN3}
    \begin{split}
    \langle v^{a,k}(p),\epsilon|J^{\hat\mu\,c}(0)|v^{b,n}(p'),\epsilon'\rangle&=\epsilon^{\hat\nu}{\epsilon'}^{\hat\rho}f^{abc}\left[\eta_{\hat\nu\hat\rho}\left(2p_{\hat\sigma}+q_{\hat\sigma}\right)+2\left(\eta_{\hat\sigma\hat\nu}q_{\hat\rho}-\eta_{\hat\rho\hat\sigma}q_{\hat\nu}\right)\right]\\
    &\times\left[\left[\eta^{\hat\mu\hat\sigma}-\frac{q^{\hat\mu}q^{\hat\sigma}}{q^2}\right]F_{v^{k}v^{n}}(q^2)+\delta_{kn}\frac{q^{\hat\mu}q^{\hat\sigma}}{q^2}\right].
    \end{split}
\end{equation}
\noindent
For the elastic case, $k=n$, the term involving $\delta_{kn}$ does not contribute due to the condition $2p\cdot q+q^2=0$. Thus, the matrix element can be written as
\noindent
\begin{equation}\label{Eq:MatrizElementNElastic}
    \begin{split}
    \langle v^{a,k}(p)&,\epsilon|J^{\hat\mu\,c}(0)|v^{b,n}(p'),\epsilon'\rangle\\
    =&f^{abc} \epsilon^{\hat\nu}{\epsilon'}^{\hat\rho}\left[\eta_{\hat\nu\hat\rho}\left(2p_{\hat\sigma}+q_{\hat\sigma}\right)+2\left(\eta_{\hat\sigma\hat\nu}q_{\hat\rho}-\eta_{\hat\rho\hat\sigma}q_{\hat\nu}\right)\right]
    \times\left[\eta^{\hat\mu\hat\sigma}-\frac{q^{\hat\mu}q^{\hat\sigma}}{q^2}\right]F_{v^{k}v^{n}}(q^2).
    \end{split}
\end{equation}
\noindent

\section{Electromagnetic form factor of vector mesons in soft-wall model}\label{vecsoftwall}

In this Appendix, we will briefly describe the electromagnetic form factors of vector mesons in the soft-wall model based on \cite{Grigoryan:2007my}. This model achieves a successful description of the Regge trajectories of vector mesons \cite{Karch:2006pv}. The electromagnetic form factor was investigate in this model considering charm mesons with chiral symmetry breaking \cite{Ahmed:2023zkk}. As presented in subsection \ref{subsec:vectormesons}, the bulk to boundary propagator obeys the differential equation \eqref{bulk to boundary propagator}

\begin{align}
     \left[\left(\partial_{u} + A_{s}^{\prime} - \Phi^{\prime}\right)\partial_{u} - q^{2}\right]V(u, q^{2}) = 0\,.
\end{align}

The soft-wall model can be obtained from the Einstein-dilaton models discussed in this work, considering warp factor and dilaton field as $A_{s} = - \ln u$ and $\Phi (u) = u^{2}$. Under these considerations, the equation above reduces to 

\begin{align}
    \left[u^{2}\partial_{u}^{2} - \left(1 + 2 u^{2}\right)u\partial_{u}\right]V(u,q^{2}) - q^{2}u^{2} V(u, q^{2}) = 0\,,\label{softdifferentialeq}
\end{align}
whose analytical solution is given by
\begin{align}
    V(q^{2}, u) = c_{1}\,u^{2} M\left(1 + \dfrac{q^{2}}{4}, 2, u^{2}\right)  + c_{2}\, u^{2} U\left(1 +  \dfrac{q^{2}}{4}, 2, u^{2}\right),\label{softsolution} 
\end{align}
where the special functions $M(a, b, x)$ and $U(a, b, x)$ are Kummer and Tricomi hypergeometric functions, respectively. We take $c_{1} = 0$, since the Kummer function is singular at $u = 0$ and the condition $V(q^2,0)=1$ leads to $c_2 = \frac{q^2}{4} \Gamma (q^2/4)$. 

The normalizable solution can be obtained from the first argument of the Tricomi function, by requiring that $1 + q^{2}/4 = - n$, considering non-negative integer values for $n$. This requirement leads to the spectrum of the $\rho$ vector mesons given by
\begin{align}
    m_{v^{m}}^{2} = 4\Lambda^{2}(n + 1) \quad n = 0,1, \dots \,.\label{vectorspectrumsoftwall}
\end{align}
The tricomi hypergeometric function admits the following integral representation for the bulk to boundary propagator
\begin{align}
    V(q^{2}, u) = -\dfrac{q^{2}}{4}\Gamma\left(\frac{q^2}{4}\right)\int_{0}^{1}\,dy\, y^{\frac{q^{2}}{4} - 1}\, \exp\left[- \dfrac{y}{1 - y}u^{2}\right]\,.\label{btbp}
\end{align}
Using an integral representation for the Tricomi function the bulk to boundary propagator
takes the form
\begin{align}
    V(q^{2}, u) = u^{2}\,\Gamma\left(\frac{q^2}{4}\right)\int_{0}^{1}\, dy\,\dfrac{y^{\frac{q^{2}}{4}}}{(1 - y)^{2}}\,\exp\left[ - \dfrac{y}{1 - y}u^{2}\right]\,.\label{rebtbp}
\end{align}
For $q^{2} = 0$, the bulk to boundary propagator reduces to $V(0, u) = 1.$

The integrand of the integral representation of the bulk to boundary propagator is related to the Laguerre polynomials by
\begin{align}
    \dfrac{y^{\frac{q^{2}}{4}}}{(1 - y)^{2}}\,\exp\left[ - \dfrac{y}{1 - y}u^{2}\right] =\sum_{m = 0}^{\infty} L_{m}^{1}(u^{2})y^{m}\,.\label{Laguerre}
\end{align}
Plugging \eqref{Laguerre} into \eqref{rebtbp} and solving the integral, we obtain
\begin{align}
    V(q^{2}, u) = 4 u^{2}\sum_{m = 0}^{\infty}\dfrac{L_{m}^{1}(u^{2})}{ 4(m + 1) + q^{2}}\,.
\end{align}
As we saw in section \ref{Sec:FormFactors}, the transition written in the vector dominance representation \eqref{vecformfactorgeneral}, adapted for the soft-wall model, becomes
\begin{align}
     F_{v^n v^k}(q^{2}) =\int_{0}^{\infty} du\, \dfrac{e^{- u^{2}}}{u}\,V(q^{2}, u)\,v^{n}(u)\,v^{k}(u)\,.\label{vecformfactorgeneralv2}
\end{align}
The modes $v^{n}(u)$ can be written in terms of Laguerre polynomials as
\begin{align}
        v^{n}(u) = u^{2}\sqrt{\dfrac{2}{n + 1}}L_{n}^{1}(u^{2})\,.\label{modesLaguerre}
\end{align}
Substituting \eqref{modesLaguerre} into \eqref{vecformfactorgeneralv2} we obtain

\begin{equation}
F_{v^{n}v^{k}}(q^{2}) = \sum_{m = 0}^{\infty} \tilde{F}_{v^{m}}\,\dfrac{\tilde{g}_{v^{m}v^{n}v^{k}}}{q^{2} + m_{v^{m}}^{2}}\,,\label{fnklaguerres}
\end{equation}
where the decay constant is given by $\tilde{F}_{v^{m}} =\Lambda^{2}\sqrt{8(m + 1)}/g_{5}$, and the couplings by
\begin{equation}
\tilde{g}_{v^m v^n v^k}
\equiv \dfrac{2\sqrt{2}\,g_{5}}{\sqrt{(m + 1)(n + 1)(k + 1)}} 
\int_{0}^{\infty} du\, u^{5} e^{-u^{2}}\,
L_m^{1}(u^{2})\, L_n^{1}(u^{2})\, L_k^{1}(u^{2}).\label{laguerres3}
\end{equation}
The result \eqref{laguerres3} provides the most general expression for the couplings among three vector mesons in the soft-wall model. The Laguerre polynomials satisfy the Rodrigues' formula \cite{Gradshteyn:1943cpj}
\begin{align}
    L_n^{\alpha}(x) = \sum_{m=0}^{n}
(-1)^m
\binom{n+\alpha}{n-m}
\frac{x^m}{m!}\,,\label{rodrigues}
\end{align}
where for our case $\alpha = 1$. Plugging \eqref{rodrigues} into \eqref{laguerres3}, we can obtain the couplings between the vector mesons displayed in tables \ref{table:VectorMesonCouplings} and \ref{table:VectorMesonCouplingsnonelastic}. The couplings involve integrals of the type
\begin{align}
    \int_{0}^{\infty} u^{2n + 1} e^{- u^{2}}\, du =\dfrac{1}{2}\Gamma(n + 1)\,.
\end{align}

We are interested in the case  $\tilde{g}_{v^m v^0 v^k}$. In this case, the result \eqref{laguerres3} reduces to
\begin{equation}
\tilde{g}_{v^m v^0 v^k}= \dfrac{2\sqrt{2}\,g_{5}}{\sqrt{(m + 1)(k + 1)}} 
\int_{0}^{\infty} du\, u^{5} e^{-u^{2}}\,
L_m^{1}(u^{2})\, L_k^{1}(u^{2}),\label{laguerres2}
\end{equation}
where $L_{0}^{1}(u^{2}) = 1$. It's convenient to rewrite the equation \eqref{laguerres2} as
\begin{align}
    \tilde{g}_{v^m v^0 v^k}= \dfrac{g_{5}\sqrt{2}}{\sqrt{(m + 1)(k + 1)}}\int_{0}^{\infty} dw\,w^{2} e^{-w}\,
L_m^{1}(w)\, L_k^{1}(w)\,,\label{laguerres2rewrite}
\end{align}
where we used $w = u^{2}$. The integral \eqref{laguerres2rewrite}, involving two Laguerre polynomials, can be written in the form \cite{Gradshteyn:1943cpj}
\begin{align}
 \int_{0}^{\infty} du\, w^{2}\,e^{-w}\, L_m^{1}(w)\, L_k^{1}(w) = (-1)^{m + k}\,2!\begin{pmatrix}
        m + 1\\
       k
   \end{pmatrix}\begin{pmatrix}
       k + 1\\
       m
   \end{pmatrix}\,.\label{integral2laguerres}
\end{align}
Plugging \eqref{integral2laguerres} into \eqref{laguerres2rewrite}
\begin{align}
    \tilde{g}_{v^m v^0 v^k}=2\,g_{5}\sqrt{2}\,(-1)^{m + k}\dfrac{\sqrt{(m + 1)(k + 1)}}{\Gamma(m - k + 2)\Gamma(k - m + 2)}\,,\label{couplingsgamma}
\end{align}
Note that $|m - k| \le 1$ in order to obtain non-vanishing strong couplings $\tilde{g}_{v^m v^0 v^k}$. This constraint for the couplings can be seen as a selection rule. Using the \eqref{couplingsgamma}, we can obtain the couplings between vector mesons, for example,
\begin{align}
     \tilde{g}_{v^0 v^0 v^1} &= - 4\pi\,,\\
      \tilde{g}_{v^1 v^0 v^1} &= 8\pi\sqrt{2}\,,\\
       \tilde{g}_{v^2 v^0 v^1} &= - 4\pi\sqrt{3}\,,
\end{align}
where we consider $k = 1$ fixed. We are interested in the case $F_{v^0 v^k}(q^{2})$. In this case \eqref{fnklaguerres}, reduces to
\begin{align}
 F_{v^0 v^k}(q^2) = 4
\sum_{m=0}^{k + 1}
\frac{\tilde{g}_{v^m v^0 v^k}}{q^{2} + 4(m + 1)}.\label{f0kvec}   
\end{align}
 We are interested in the case $F_{v^0 v^k}(q^{2})$. Using the couplings \eqref{laguerres2rewrite}, we can describe the transition form factor of vector mesons between the ground state, $v^{0}$, and any $v^k$-state, in which can be written as
 \begin{align}
     F_{v^0 v^k}(q^{2}) = \sum_{m = 0}^{\infty} \dfrac{ \tilde{g}_{v^m v^0 v^k}}{m_{v^{m}}^{2} + q^{2}}\,.\label{f0k}
 \end{align}
 From the general result \eqref{f0k}, we can obtain particular cases. For example, the transition form factor of vector mesons between the ground and first excited states, $F_{v^0 v^1}(q^{2})$, $F_{v^0 v^2}(q^{2})$ and $F_{v^0 v^3}(q^{2})$. From \eqref{f0k}, we obtain
\begin{align}
    F_{v^0 v^1}(q^{2}) &= \sqrt{2}\,\left(-\dfrac{1}{1 + \dfrac{q^{2}}{m_{v^{0}}^{2}}} + \dfrac{2}{1 + \dfrac{q^{2}}{m_{v^{1}}^{2}}} - \dfrac{1}{1 + \dfrac{q^{2}}{m_{v^{2}}^{2}}}\right)\,,\label{Eq:F01}\\
     F_{v^0 v^2}(q^{2}) &= \sqrt{3}\,\left(-\dfrac{1}{1 + \dfrac{q^{2}}{m_{v^{1}}^{2}}} + \dfrac{2}{1 + \dfrac{q^{2}}{m_{v^{2}}^{2}}} - \dfrac{1}{1 + \dfrac{q^{2}}{m_{v^{3}}^{2}}}\right)\,,\label{Eq:F02}\\
      F_{v^0 v^3}(q^{2}) &= -\dfrac{2}{1 + \dfrac{q^{2}}{m_{v^{2}}}} + \dfrac{3}{1 + \dfrac{q^{2}}{m_{v^{3}}}} -\dfrac{2}{1 + \dfrac{q^{2}}{m_{v^{4}}}}\,, 
\end{align}
where $m_{v^{k}}^{2} = 4\,\Lambda^{2}(k+1),\, k=0,1,\cdots$.  Note that $F_{v^0 v^1}(q^{2})$ receives contributions from the ground state, the first excited state, and the second excited state, while all higher contributions vanish due to the selection rule. 
In contrast, the transition form factor $F_{v^0 v^2}(q^{2})$ does not receive a contribution from the ground state and instead involves only the first, second, and third excited states. 
Finally, the transition form factor $F_{v^0 v^3}(q^{2})$ receives contributions from the second, third, and fourth excited states. Another interesting example is to consider only the ground state, $F_{v^0 v^0}(q^{2})$. By considering $k = 0$, the result \eqref{f0k} leads to us
\begin{align}
      F_{v^0 v^0}(q^{2}) =\dfrac{2}{1 + \dfrac{q^{2}}{m^{2}_{v^{0}}}} - \dfrac{1}{1 + \dfrac{q^{2}}{m^{2}_{v^{1}}}}\,. \label{formfactoranalyticalSW}
\end{align}
This result \eqref{formfactoranalyticalSW} was obtained in \cite{Grigoryan:2007my}. Note that the form factor in the soft-wall model \eqref{formfactoranalyticalSW} is dominated by the ground state and first excited state, since all higher contributions vanish. This result is consistent with the couplings listed in table \ref{table:VectorMesonCouplings}, where the first two states provide the dominant contributions.

\begin{figure}[htb!]%
    \centering
    \begin{subfigure}{0.45\textwidth}
        \centering
        \includegraphics[width=\textwidth]{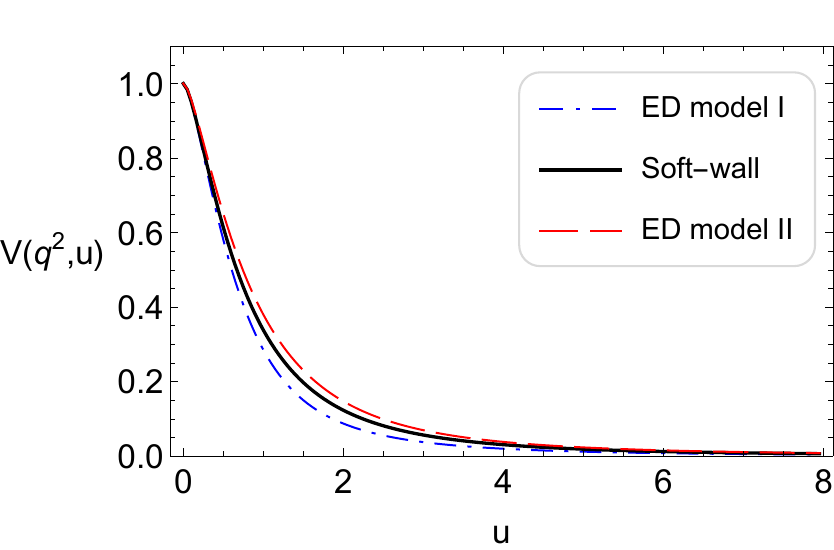}
    \end{subfigure}
    \hspace{0.05\textwidth}
    \begin{subfigure}{0.45\textwidth}
        \centering
        \includegraphics[width=\textwidth]{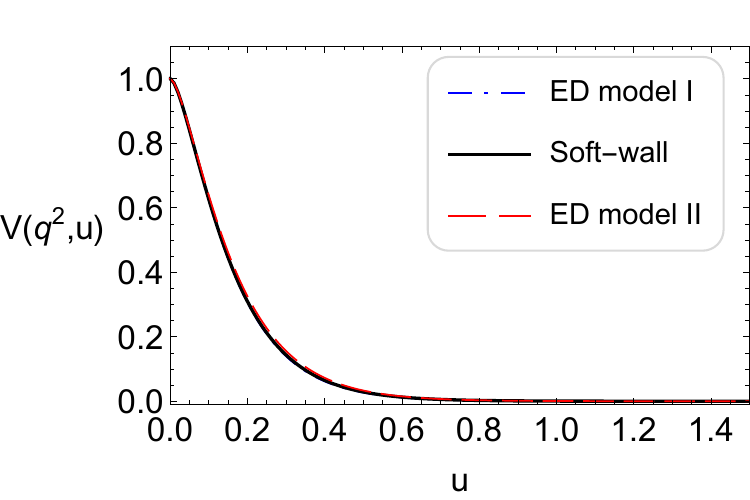}
    \end{subfigure}
    \caption{{\bf Left panel:} Bulk to boundary propagator for the Einstein-dilaton models I and II and soft wall model. All curves where obtained considering the dimensionless parameter $q^2/\Lambda^2=5$. {\bf Right panel:} Bulk to boundary propagator in the Einstein-dilaton models I and II and soft wall model. All curves where obtained considering the dimensionless parameter $q^2/\Lambda^2=90$.}
    \label{Plot:BtBProp}
\end{figure}

\section{Bulk to boundary propagator}\label{vecbtbp}

In this Appendix we investigate the solution of the differential equation \eqref{bulk to boundary propagator} satisfied by the bulk to boundary propagator $V(q^2,u)$,
\begin{equation}\label{Eq:Bulk-to-Boundary-Prop}
    V''(q^2,u)+(A_s'-\Phi')V'(q^2,u)-q^2V(q^2,u)=0.
\end{equation}
Different from the soft-wall model, where the solution of the differential equation is analytic \cite{Grigoryan:2007my}, there are no analytic solutions for models I and II. Then, we need to solve numerically the differential equation. For doing that, we need the asymptotic solution either, close to the boundary or far from the boundary, depending on the strategy followed to solve the differential equation; here we integrate the problem from the IR region to the boundary, thus, we need the asymptotic solution far from the boundary which is obtained considering the ansatz 
\begin{equation}\label{Eq:SolBtBPIR}
    V(q^2,u)=u^{-\frac{q^2}{2}}\left(c_0+\frac{c_1}{u}+\frac{c_2}{u^2}+\frac{c_3}{u^3}+\frac{c_4}{u^4}+\cdots\right).\quad u\to\infty
\end{equation}
The coefficients: $c_1,c_2,\cdots$, are obtained by plugging \eqref{Eq:SolBtBPIR} in \eqref{Eq:Bulk-to-Boundary-Prop} and solving order-by-order the resulting equation. The odd coefficients are zero for the soft-wall and Einstein-dilaton  models I and II, while the first even for soft-wall model are:
\noindent
\begin{equation}\label{Eq:CoeffIRSW}
    c_2=-\frac{q^2}{16}(4+q^2)c_0,\quad c_4=\frac{q^2}{512}\left(4+q^2\right)^2(8+q^2)c_0,
\end{equation}
\noindent
the coefficients for model I are:
\noindent
\begin{equation}\label{Eq:CoeffIRMI}
    c_2=-\frac{q^2}{16}(1+q^2)c_0,\quad c_4=\frac{q^2}{512}\left(29(1+q^2)+q^4(10+q^2)\right)c_0,
\end{equation}
\noindent
while for model II are:
\noindent
\begin{equation}\label{Eq:CoeffIRMII}
    c_2=-\frac{q^2}{32}(11+2q^2)c_0,\quad c_4=\frac{q^2}{2048}\left(10+q^2\right)\left(89+36 q^2+4q^4\right)c_0.
\end{equation}

\noindent
We fix the constant in the IR to $c_0=1$, then, integrate numerically up to an UV cut-off $\epsilon$. This procedure does not guarantee that the UV boundary condition $V(q^2,0)=1$ is satisfied. In turn, we can use the fact that the differential equation is linear and obtain the solution satisfying the UV boundary condition rescaling the numerical solution by the value of the bulk to boundary propagator at the cut-off, i.e., $V(q^2,u)=\widetilde{V}(q^2,u)/\widetilde{V}(q^2,\epsilon)$, where $\widetilde{V}(q^2,u)$ is the numerical solution of \eqref{Eq:Bulk-to-Boundary-Prop} that does not satisfy the boundary condition. It is worth to know the behavior of the bulk to boundary propagator for different values of $q$, specially for small-$q^2$ and large-$q^2$. Thus, our numerical results for small-$q^2$ are displayed in left panel of Fig.~\ref{Plot:BtBProp}, where continuous line represents the results of the soft-wall model, dot-dashed blue line of model I, and dashed red line of model II. In turn, the numerical results for the bulk to boundary propagator in the limit of large-$q^2$ are displayed in the right panel of Fig.~\ref{Plot:BtBProp}. These results suggest us that the relevant contribution of the bulk to boundary propagator is restricted to the region close to the boundary in the limit of large-$q^2$, as highlighted in Ref.~\cite{Grigoryan:2007my}.

\begin{figure}[htb!]
    \centering
    \begin{subfigure}{0.45\textwidth}
        \centering
        \includegraphics[width=\textwidth]{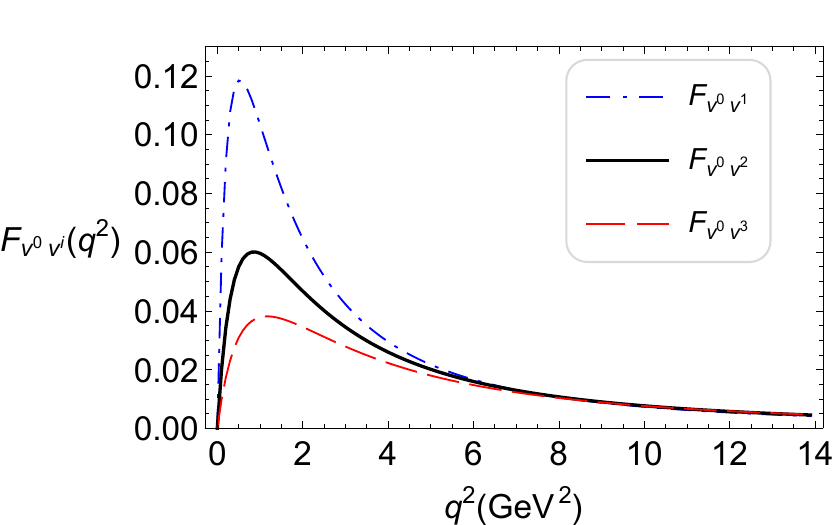}
    \end{subfigure}
    \hspace{0.05\textwidth}
    \begin{subfigure}{0.45\textwidth}
        \centering
        \includegraphics[width=\textwidth]{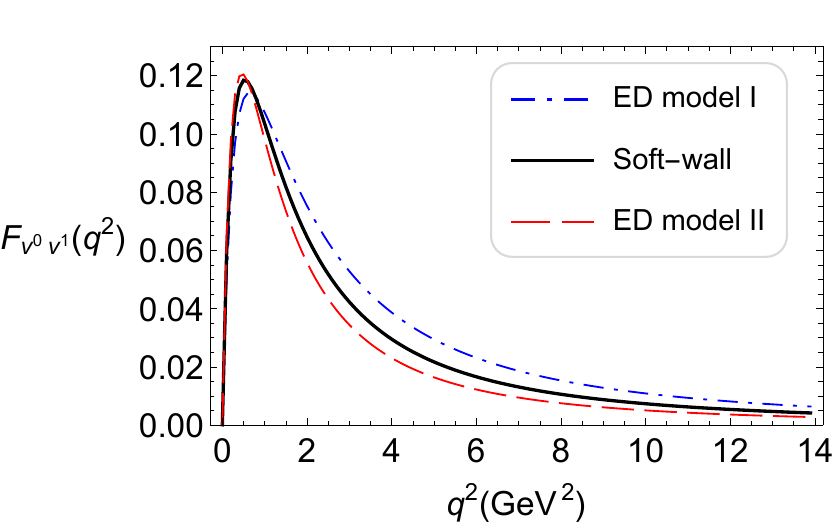}
    \end{subfigure}
    \caption{{\bf Left panel}: Non-elastic form factors of vector mesons for the soft-wall model. {\bf Right panel}: Comparison of non-elastic form factor $F_{v^0v^1}$ of vector mesons for model I (dot-dashed blue), soft-wall (solid black line), and model II (dashed red line).}
    \label{Plot:FFSW}
\end{figure}

Once the solution of the bulk to boundary propagator is available we can integrate numerically the expression for the generalized form factors, given by Eq.~\eqref{vecformfactorgeneral}, to get the elastic form factor $F_{v^{n}v^{n}}(q^2)$ and the non-elastic form factors $F_{v^{n}v^{k}}(q^2)$. Our numerical results of the non-elastic form factors for the soft-wall model are displayed in the left panel of Fig.~\ref{Plot:FFSW}. Meanwhile, right panel of the same figure shows a comparison of the non-elastic form factor $F_{v^0v^1}$ for model I (dot-dashed blue line), soft-wall model (solid black line), and model II (dashed red line).

\bibliographystyle{utphys}

\bibliography{FormFactors}

\end{document}